\documentclass[letter,preprint,aps,superscriptaddress,nofootinbib,tightenlines,preprintnumbers,cites]{revtex4}
\usepackage{graphicx}
\usepackage{enumerate}
\usepackage{bm}
\usepackage{slashed}
\usepackage{multirow}
\usepackage{verbatim}
\usepackage{amsmath,amssymb}
\usepackage{epsfig}
\usepackage{hhline}
\usepackage{placeins}
\newcommand\mpi {m_{\pi}}

\newcommand\Eq[1]{Eq.~(\ref{eq:#1})}
\newcommand\Eqs[2]{Eqs.~(\ref{eq:#1})-(\ref{eq:#2})}
\newcommand\Fig[1]{Fig.~\ref{fig:#1}}
\newcommand\Figtwo[2]{Figs.~\ref{fig:#1} and \ref{fig:#2}}
\newcommand\Figs[2]{Figs.~\ref{fig:#1}-\ref{fig:#2}}
\newcommand\Sec[1]{Sec.~\ref{sec:#1}}

\newcommand\Tab[1]{Table~\ref{tab:#1}}
\newcommand{\be}{\begin{equation}}
\newcommand{\ee}{\end{equation}}
\newcommand\beq{\begin{eqnarray}}
\newcommand\eeq{\end{eqnarray}}

\newcommand{\Tr}{{\rm Tr\,}}

\newcommand{\mybar}[1]%
        {\kern 0.6pt\overline{\kern -0.6pt#1\kern -0.6pt}\kern 0.6pt}
\begin{document}

\preprint{MIT-CTP-4491}
\preprint{UM-DOE/ER/40762-529}

\title{Baryon masses at nonzero isospin/kaon density}

\author{William Detmold}
\email{wdetmold@mit.edu}
\affiliation{Center for Theoretical Physics, Massachusetts Institute of Technology, Cambridge, MA 02139, USA}

\author{Amy N. Nicholson}
 \email{amynn@umd.edu} 
\affiliation{Maryland Center for Fundamental Physics, Department of Physics,
University of Maryland, College Park MD 20742-4111, USA}


 \date{\today}
 
 \begin{abstract}
We present a lattice QCD calculation of the ground-state energy shifts of various baryons in a medium of pions or kaons at a single value of the quark mass corresponding to a pion mass of $m_{\pi}\sim 390$ MeV and a kaon mass of $m_{K}\sim 540$ MeV, and in a spatial volume $V\sim (4\mathrm{fm})^3$. All systems are created using a canonical formalism in which quark propagators are contracted into correlation functions of fixed isospin/kaon density. We study four different systems, $\Sigma^{+}(\pi^{+})^n$, $\Xi^0(\pi^{+})^n$, $p(K^{+})^n$, and $n(K^{+})^n$, for up to $n=11$ mesons. From the ground-state energy shifts we extract two- and three-body scattering parameters, as well as linear combinations of low-energy constants appearing in tree-level chiral perturbation theory.
 
\end{abstract}

\maketitle

\section{Introduction}

The field of nuclear physics is largely defined by the study of the multi-hadron systems that compose much of the universe that we observe. Understanding how the diverse array of multi-hadron systems emerges from the deceptively simple underlying theory defining their interactions (QCD) has become a central goal for the field. However, because of the non-perturbative nature of QCD in the low-energy regime, the only known method that allows first-principles calculations is lattice QCD. Because of the well known signal-to-noise problem \cite{Lepage:1989hd}, lattice QCD calculations of correlation functions involving multiple baryon systems are severely hindered, while the sign problem prohibits calculations with non-zero baryon chemical potential \cite{Barbour:1986jf,Kogut:1994eq}. Thus, lattice QCD calculations of multiple hadron systems have been largely limited to studies of mesons. Recent results involving multiple pions \cite{Detmold:2008fn,Beane:2007es,Detmold:2008bw,Detmold:2012wc,Detmold:2012pi}, multiple kaons \cite{Detmold:2008yn}, and mixed pion-kaon systems \cite{Detmold:2011kw}, have provided insight into the new problems faced when dealing with multiple particles in lattice calculations and highlighted the importance of many-body interactions for even moderate densities. More recently, calculations of multibaryon systems have begun \cite{Beane:2009gs,Yamazaki:2009ua,Yamazaki:2012hi,Beane:2012vq,Detmold:2012eu}.

In addition to providing a test bed for lattice calculations of many-body systems, multiple meson systems are of particular interest as they may provide crucial information about the equation of state of neutron stars and the late time evolution of heavy ion collisions. As interacting bosonic gases, they also allow us to study the phenomenon of Bose-Einstein condensation. Pion condensation \cite{Migdal,Sawyer:1972cq,Scalapino:1972fu,PhysRevLett.30.1340,Migdal:1979je} and kaon condensation \cite{Kaplan:1986yq,Kaplan:1987sc} may be of phenomenological relevance as they have been proposed to occur in the interior of neutron stars. In this realm, lattice QCD calculations could have a strong impact in the field of astrophysics. The existence of a kaon condensate, for example, relies heavily on the poorly known ``strangeness'' content of the nucleon and on $KN$, $KNN$, $\dots$ interactions.

As a first step toward calculations of many-body systems involving both baryons and mesons, in this work we calculate properties of systems with up to 11 mesons and a single baryon. We first determine the ground state energies from correlation functions of appropriately contracted quark propagators. The systems are chosen to avoid annihilation diagrams between valence quarks, and are listed in Tab.~\ref{tab:barpisys}. While perhaps the most phenomenologically interesting, pion-nucleon and $K^{-}$-nucleon systems are absent from this study as they are computationally prohibitive to compute given our current resources. 

From these energies, we extract meson-baryon scattering lengths and present the first calculation of three-body interactions between mesons and baryons. Finally, we use leading order results from heavy baryon chiral perturbation theory (HB$\chi$PT) at finite isospin and/or strangeness chemical potential to extract low-energy constants (LECs) related to the pionic or kaonic medium contributions to the baryon masses.

\begin{table}
\centering
\caption[The simplest hadronic content of baryon-meson systems presented in this work.]{\label{tab:barpisys}The simplest hadronic content of baryon-meson systems presented in this work. For each system, we have chosen a single baryon and $n=1-11$ mesons. The valence quark content, isospin, and strangeness are also shown. These particular systems were chosen so as to avoid possible annihilation between valence quarks.}
\begin{tabular}{|c|c|c|c|}
\hline
System & Quark Content & I & S\\
\hline
$\Xi^0 (\pi^{+})^n$ & uss(u$\bar{\mbox{d}})^n$ & $n+\frac{1}{2}$ & -2\\
$\Sigma^{+} (\pi^{+})^n$ & uus(u$\bar{\mbox{d}})^n$ & $n+1$ & -1\\
$p (K^{+})^n$ & uud(u$\bar{\mbox{s}})^n$ & $\frac{n}{2} +1$& $n$\\
$n (K^{+})^n$ & udd(u$\bar{\mbox{s}})^n$ &$ \frac{n}{2}+\frac{1}{2}$ & $n$\\
\hline
\end{tabular}
\end{table}

\section{Baryons in a meson condensate from HB$\chi$PT}
One may investigate the phases of QCD at finite isospin chemical potential, $\mu_I \equiv \frac{1}{2} (\mu_u-\mu_d)$, using two flavor Chiral Perturbation Theory ($\chi$PT) \cite{Son:2000xc,Son:2000by}. It is expected that at zero temperature there is a phase transition at $\mu_I = \mpi/2$ to a Bose-condensed phase. This transition can be seen by parametrizing the chiral condensate as
\beq
\langle \bar{q}q\rangle = \left( \begin{array}{cc}
\cos \alpha & -\sin \alpha \\
\sin \alpha & \cos\alpha \\
\end{array} \right) \ .
\eeq
In vacuum, $\cos \alpha = 1$, giving the usual chiral condensate, however in a BEC state, minimization of the leading order effective potential results in $\cos \alpha = \mpi^2/\mu_I^2$. In the condensed phase, there exists one massless mode, corresponding to a linear combination of the vacuum $\pi^{+}$ and the $\pi^{-}$. At higher $\mu_I$, other modes such as a combination of $\rho^{+/-}$ may also condense. Finally, as the chemical potential is raised further, a crossover to a BCS phase consisting of weakly interacting quark-antiquark pairs is expected based on the asymptotic freedom of QCD \cite{Son:2000xc,Son:2000by}.  

The masses of several low-lying baryons have also been computed using $SU(2)$ HB$\chi$PT at nonzero $\mu_I$ \cite{Bedaque:2009yh}. The presence of the condensate mixes the baryons; in particular, the ground-states produced by an operator with the quantum numbers of the vacuum $\Sigma^{+}$ and $\Xi^0$ are a linear combination of the vacuum  $\{\Sigma^{+}\, , \, \Sigma^{-}\}$ and $\{\Xi^0\, , \, \Xi^{-}\}$, respectively. For the charged $\Sigma$ state, the mass as a function of isospin chemical potential to $\mathcal{O}(p^2)$ (tree-level) is given by
\beq
\label{eq:Sigmass}
M_{\Sigma^{+}}(\mu_I) &=& M_{\Sigma}^{(0)}+4c_1^{\Sigma} m_{\pi}^2\cos \alpha +(c_2^{\Sigma}+c_3^{\Sigma}+c_6^{\Sigma}+c_7^{\Sigma})\mu_I^2 \sin^2\alpha \cr
&-&\mu_I\sqrt{\cos^2\alpha+(c_6^{\Sigma}+c_7^{\Sigma})^2\mu_I^2 \sin^4\alpha} \ ,
\eeq
where, as discussed above, the angle $\alpha$ parametrizes the transition from vacuum to a pion condensed phase, $M_X^{(0)}$ is the mass of the baryon in the chiral and zero chemical potential limits, and all LECs, $c_{i}^{X}$, are as defined in \cite{Bedaque:2009yh}. 
Similarly, for the relevant $\Xi$ state in a pion condensate, the mass is given by \cite{Bedaque:2009yh},
\beq
M_{\Xi^{0}}(\mu_I) &=&  M_{\Xi}^{(0)}-\frac{\mu_I}{2} \cos \alpha +4c_1^{\Xi}m_{\pi}^2 \cos \alpha + \left(c_2^{\Xi}-\frac{g_{\Xi\Xi}^2}{8M_{\Xi}^{(0)}}+c_3^{\Xi}\right) \mu_I^2 \sin^2\alpha \, .
\eeq
Including the strange quark using $SU(3)$ HB$\chi$PT yields the same form for the masses, but with different  combinations of $SU(3)$ parameters. 

The QCD phase diagram including nonzero kaon chemical potential, $\mu_K = \frac{1}{2}(\mu_u-\mu_s)$, has also been investigated in Ref.~\cite{Kogut:2001id} using similar methods to those used for the isospin case. We may repeat the calculations of Ref.~\cite{Bedaque:2009yh} to obtain the masses of the nucleons as a function of the kaon chemical potential. The nucleons now mix with the hyperons when the condensate forms; in particular, the proton mixes with the $\Xi^{-}$, while the neutron mixes with the $\Sigma^{-}$. Due to approximate $SU(3)$ symmetry the mass of the baryon sharing the quantum numbers of the vacuum neutron has the same form as that for the $\Xi$ mass with an isospin chemical potential (excluding additional $\mu$-independent quark mass terms), 
\beq
M_{n}(\mu_K) &=&M_{n}^{(0)} -\frac{\mu_K}{2} \cos \alpha + \left(2b_0+b_D-b_F\right)m_{K}^2 \cos \alpha \cr
&+&\frac{1}{4}\left(b_1-b_2+b_3+b_4-b_5+b_6+2b_7+2b_8\right)\mu_K^2 \sin^2\alpha\ ,
\eeq
where the $b_i$ are LECs of the $SU(3)$ HB$\chi$PT Lagrangian.

We also find that the mass of the baryon sharing the quantum numbers of the proton as a function of $\mu_K$ is similar in form to the $\Sigma$ mass, \Eq{Sigmass}, but with an additional $SU(3)$ breaking term proportional to the strange and light quark mass difference,
\beq
\label{eq:Protmass}
M_{p}(\mu_K) &=& M_{p}^{(0)} +2 \left(b_0+b_D\right)m_{K}^2 \cos \alpha +\frac{1}{2}\left(b_1+b_3+b_4+b_6+b_7+b_8\right)\mu_K^2 \sin^2\alpha\cr
&-&\sqrt{\left(2b_F(m_K^2-m_{\pi}^2)+\mu_K\cos\alpha\right)^2+\frac{1}{4}(b_1-b_3+b_4-b_6)\mu_K^4 \sin^4\alpha} \ .
\eeq

\section{Extracting two- and three-body interactions from multi-hadron energies}

The methodology for calculating scattering phase shifts from the energy levels of two particles in a finite box was originally established in quantum field theory by L\"{u}scher \cite{Luscher:1986pf,Luscher:1990ux} (see also \cite{Hamber:1983kx}). There have been several recent works which calculate the ground-state energies of multiple bosons in a periodic box as a perturbative expansion in the size of the box ($L$) using the 2-particle phase shifts and 3-particle interaction parameter as inputs. These calculations have been performed for identical bosons \cite{Beane:2007qr,Detmold:2008gh} to $\mathcal{O}(L^{-7})$, as well as mixed species of bosons \cite{Smigielski:2008pa} to $\mathcal{O}(L^{-6})$. 

We note that so long as there is only a single fermion which carries the spin of the system, Fermi statistics do not play a role and we may use the mixed species form for the energy shift where the baryon represents one of the species and the mesons, the other. Thus, in this work, we will use the following relation for the energy shift of a single baryon of mass $m_B$ and $n$-mesons of mass $m_M$ due to their interactions:

\beq
\label{eq:LuscherExp}
    \Delta E_{MB}(n,L) &\equiv& E_{MB}(n,L) - E_{B}(L) - E_{M}(n,L)\cr
&=&\frac{2{\pi}\bar{a}_{MB}n}{\mu_{MB}L^3}\left[1-\left(\frac{\bar{a}_{MB}}{\pi{L}}\right)\mathcal{I}+\left(\frac{\bar{a}_{MB}}{\pi{L}}\right)^{2}  \right. \cr
   &&\times \left(\mathcal{I}^2 +\mathcal{J}\left[-1+2\frac{\bar{a}_{MM}}{\bar{a}_{MB}}(n-1)\left(1+\frac{\mu_{MB}}{m_M}\right)\right.\right.\cr
    &&\left.+\left(\frac{\bar{a}_{MB}}{\pi{L}}\right)^{3}\left(-\mathcal{I}^{3}
    +\sum_{i=0}^{2}\left(f^{\mathcal{I}\mathcal{J}}_{i}\mathcal{I}\mathcal{J}
    +f^{\mathcal{K}}_{i}\mathcal{K}\right)\left(\frac{\bar{a}_{MM}}{\bar{a}_{MB}}\right)^{i}\right)
    \right]\nonumber\\
    &&+\frac{n(n-1)\bar{\eta}_{3,MMB}(L)}{2L^6}+\mathcal{O}(L^{-7}) \ ,
\eeq
where $E_{MB}(n,L)$, $E_{M}(n,L)$ are the ground-state energies of the $n$ meson sytems with and without a baryon, respectively, $\bar{a}_{MB}$ and $\bar{a}_{MM}$ are the inverse of the scattering phase shifts, $\left(p \cot \delta(p)\right)^{-1}$, corresponding to meson-baryon and meson-meson interactions, respectively. For large volume, we may relate these quantities to their respective scattering lengths, $a$, and effective ranges, $r$, using the effective range expansion \cite{Detmold:2008fn},
\beq
    a&=&\bar{a}-\frac{2{\pi}\bar{a}^{3}r}{L^3}\ .
\eeq
The geometric constants are given by
\beq
\mathcal{I} = -8.9136329 \ , \qquad \mathcal{J} = 16.532316 \ , \qquad \mathcal{K} = 8.4019240 \ ,
\eeq
and we use the following notation,
\beq
f_0^{\mathcal{I}\mathcal{J}} &=& n+2 \ , \qquad f_0^{\mathcal{K}}= n-2 \ , \qquad f_1^{\mathcal{I}\mathcal{J}} = 2(1+2\frac{\mu_{MB}}{m_M})(1-n) \ , \cr
f_1^{\mathcal{K}} &=& 2 \mu_{MB} \frac{m_M^2 + 9 m_B^2 + 4 m_B m_M}{m_M m_B^2} (n-1) \ , \qquad f_2^{\mathcal{I}\mathcal{J}} = \mu_{MB} \frac{m_B - m_M}{m_B m_M} (1-n) \ ,\cr
f_2^{\mathcal{K}} &=& \frac{\mu_{MB}^2}{m_B^3 m_M^2} (1-n) (m_B^3 (13n - 45) - m_M^3 + m_M m_B^2 (14n - 39) + 5 m_M^2 m_B (n - 3)) \ .
\eeq
The volume-dependent, but renormalization group invariant, three-body interaction, $\bar{\eta}_{3,MMB}(L)$, is expected to behave logarithmically with the volume for large volumes. Its form is shown explicitly in \cite{Detmold:2008gh}.

\section{Methodology and Details of the Lattice Calculation}

\subsection{Correlation Functions}

Naively, the number of contractions of the quark propagators necessary to form a correlation function for a single baryon and $n$ mesons is $N_u!N_d!N_s!$, where $N_i$ is the number of quarks of flavor $i$ in the chosen interpolating operators, which would clearly be prohibitive for the problem at hand. Instead, we employ a method based on the formalism developed in \cite{Detmold:2008fn,Beane:2007es,Detmold:2011kw}.

Here we briefly review the method devised to perform contractions for multiple meson propagators. The mesons are first packaged into the following $12 \times 12$ matrices by contracting the quark and anti-quark propagators at the sink:
\beq
\Pi_{a, \alpha, b, \beta} &=& \sum_{c,\gamma} \sum_{\mathbf{x}}\left[ S_u(\mathbf{x},t;\mathbf{0},0) \gamma_5 \right]_{b,\beta,c,\gamma}\left[ S_d^{\dagger}(\mathbf{x},t;\mathbf{0},0) \gamma_5 \right]_{a,\alpha,c,\gamma} \to \Pi_{i,j} \cr
K_{a, \alpha, b, \beta} &=& \sum_{c,\gamma} \sum_{\mathbf{x}}\left[ S_u(\mathbf{x},t;\mathbf{0},0) \gamma_5 \right]_{b,\beta,c,\gamma}\left[ S_s^{\dagger}(\mathbf{x},t;\mathbf{0},0) \gamma_5 \right]_{a,\alpha,c,\gamma} \to K_{i,j} \, ,
\eeq
where $S_q(\mathbf{x},t;\mathbf{0},0)$ is the propagator for quark flavor $q$ from point $(\mathbf{0},0)$ to $(\mathbf{x},t)$, $(a,b,c)$ indices represent color, Greek indices represent spin, and the indices $(i,j)$ run over the 12 color/spin combinations. The sum over $\mathbf{x}$ projects the meson onto zero momentum.

As shown in Ref.~\cite{Detmold:2011kw}, correlation functions, $C_{n,m}(t)$, for $n$ pions and $m$ kaons can be formed using the following relation:
\beq
\label{eq:DetExp}
\det(1 + \lambda \Pi + \kappa K)=\frac{1}{12!}\sum_{j=1}^{12}\sum_{k=0}^{j}\left( \begin{array}{c}
12 \\
j \\
\end{array} \right)\left( \begin{array}{c}
j \\
k \\
\end{array} \right) \lambda^k \kappa^{j-k} C_{k,j-k}(t) = e^{\Tr[\log(1 + \lambda \Pi + \kappa K)]} \, .
\eeq
By expanding the right-hand side of this equation and collecting terms with $n$ powers of $\lambda$ and $m$ powers of $\kappa$, one finds the desired correlation function in terms of products of traces of matrix products. Note that this method limits the study to systems involving up to 12 quarks of a given flavor due to the Pauli exclusion principle. Methods for larger systems are discussed in \cite{Detmold:2010au,Detmold:2012wc}.

To include baryon fields, we begin with the following baryon ``block'':
\beq
B_{a,\alpha, b,\beta, c,\gamma,\lambda}= \sum_{\sigma,h,i,j}[S_{q_1}C\gamma_5]_{a,\alpha,h,\sigma}[S_{q_2}]_{b,\beta,i,\sigma}[S_{q_3}]_{c,\gamma,j,\lambda}\epsilon_{h,i,j} \, ,
\eeq
where $q_{1,2,3}$ represents the flavors of the three quarks, $C$ is the charge conjugation matrix, and spatial coordinates have been suppressed (all propagators are from $(\mathbf{0},0)$ to $(\mathbf{x},t)$, and $\mathbf{x}$ is summed over as in the mesonic case). The remaining spin indices must be contracted using the proper parity projector depending on the desired state. 

To perform the contractions for the positive parity $\Xi$ system, we partially contract the s-quark indices as follows:
\beq
B_{a,\alpha,d,\rho}^{(\Xi)}= \sum_{b,\beta,c,\gamma,\lambda} B_{a,\alpha, b,\beta, c,\gamma,\lambda} \epsilon_{d,b,c}(C\gamma_5)_{\beta,\rho}(1\pm \gamma_4)_{\gamma,\lambda}\to B_{i,j}^{(\Xi)} \, .
\eeq
With this form, the s quarks combine into a $\bar{3}$ color irrep and leave a single spin index open, effectively behaving as the antiquark in the meson blocks. Using this, we can compose correlation functions using the multiple species formalism, \Eq{DetExp}, where we replace $K\to B^{(\Xi)}$ and collect terms involving only a single baryon ($k=1$). A neutron block can be similarly constructed and contracted with a kaon medium.

The contractions for the $\Sigma$ and proton are more complicated because each of the doubly represented up quarks must be antisymmetrized with the corresponding up quarks in the mesons. The solution is to embed the baryon into a higher-dimension matrix as follows:
\beq
B_{d,\alpha,b,\rho,f,\lambda,c,\gamma}^{(\Sigma,p)}= \sum_{a,\beta,\sigma} B_{a,\alpha, b,\beta, c,\gamma,\lambda} \epsilon_{a,d,f}(C\gamma_5)_{\beta,\rho}(1\pm \gamma_4)_{\sigma,\lambda}\to B_{I,J}^{(\Sigma,p)} \, .
\eeq
This is now a $144 \times 144$ matrix, with indices $I,J$ running over the spin/color indices of both up quarks. Corresponding $144 \times 144$ matrices for the mesons can be formed by taking an outer product of a meson block with the $12\times 12$ identity matrix, $\{\Pi \otimes 1\ ,\ 1\otimes \Pi\}$, and likewise for the kaons. We now use \Eq{DetExp}, replacing $K\to B^{(\Sigma,p)}$ ($\Pi\to B^{(\Sigma,p)}$) and $\Pi \to \Pi \otimes 1+1\otimes \Pi$ ($K \to K \otimes 1+1\otimes K$) for a single baryon in a pion (kaon) medium. 

Using this method, the computational cost is greatly reduced by converting the index contractions into traces of matrix products, for which we may use highly optimized linear algebra packages. Furthermore, because we perform the contractions for all systems with up to twelve mesons simultaneously, we only need to calculate and store all possible traces of matrix products once. These may then be combined according to the expansion of \Eq{DetExp}. For example, if we wish to calculate the $n$th meson correlator and have already computed that for $n-1$ mesons, we are only required to calculate a single term, $\Tr \left[ B \Pi^{n} \right]$, as all lower powers have already been computed and the matrix product $B\Pi^{n-1}$ is also known.

\subsection{Gauge field configurations and quark propagators}
To perform the following calculation, we use gauge configurations computed by the Hadron Spectrum Collaboration (for details, see Ref.~\cite{Lin:2008pr}). The gauge fields were created using a $n_f=2+1$-flavor anisotropic dynamical tadpole-improved clover fermion action \cite{Chen:2000ej} with a Symanzik-improved gauge action \cite{Symanzik:1983dc,Symanzik:1983gh,Luscher:1984xn,Luscher:1985zq}. The ensembles have a spatial lattice spacing of $b_s=0.1227(8)$ fm, a pion mass of $m_{\pi} \sim 386$ MeV, and a kaon mass of $m_K \sim 543$ MeV. The renormalized anisotropy parameter, $\xi=b_{t}/b_s = 3.469(11)$, was determined in Ref.~\cite{Beane:2011sc}. We use configurations on a large volume ($32^3$) to ensure that we are near threshold in our extraction of scattering parameters, and a large temporal extent ($T=256$) to eliminate thermal effects in our correlation functions. We use the quark propagators from Ref.~\cite{Beane:2009ky}, which were generated using the same fermion action as was used for gauge field generation (for more details, see Ref.~\cite{Beane:2009ky}). 

\subsection{Analysis}
This calculation uses approximately 200 gauge configurations with 150 sources on each configuration, calculated at randomly chosen positions throughout the lattice. The correlators from all sources on a given gauge configuration are first averaged; the resulting set is then resampled using the bootstrap method to enable the calculation of uncertainties. To determine the energy splittings, one may calculate the following ratio for each bootstrap ensemble,
\beq
\Delta M_{\mathrm{eff}}^{(n)}(t)=\ln \left(\frac{C_{B,n}(t)/C_{B,n}(t+1)}{[C_B(t)/C_B(t+1)][C_{n}(t)/C_{n}(t+1)]}\right)\, ,
\label{eq:massshift}
\eeq
where $n$ is the number of mesons in the system and $B$ indicates the presence of a baryon. In the limit of large Euclidean time, this quantity gives the energy shift to the baryon mass resulting from its interactions with the mesons. To improve the statistics of our calculation, we compute the correlation function for both parities and average the results after performing the proper time reversal. 
We fit the energies using a correlated $\chi^2$ method for various time intervals using a fit function taking into account thermal effects \cite{Detmold:2011kw}, as well as a simpler constant plus exponential fit to account for contamination from excited states. It was determined that for the energy splittings on these ensemble, thermal states have negligible effects and the best fit is given by a constant plus exponential fitting function for late times, however, thermal effects will be important on ensembles with a shorter temporal extent. In all cases, a plateau region of several time steps is observed and confirmed to agree with the constant plus exponential fit, indicating that both thermal and excited state contributions are sufficiently suppressed. Statistical uncertainties are calculated using the bootstrap method, and a fitting systematic uncertainty is found by varying the fitting region within our best fit region, as well as allowing the endpoints to vary by $\Delta \tau = \pm 2$.

To determine the two- and three-body parameters, as well as the LECs, bootstrap ensembles of fits to the energy splittings are created for the best fit time intervals, as well as for all time intervals beginning and ending within $\pm 2$ of the best fit. These ensembles are then used to perform correlated $\chi^2$ fits to the appropriate expressions, \Eq{LuscherExp}, to extract the interaction parameters, and a fitting systematic uncertainty is determined using the ensembles for different time intervals. In some cases, the statistics of the extraction can be improved by performing simultaneous fits to both the baryon plus mesons systems and the pure many mesons systems. Finally, the effects of the uncertainty in the anisotropy parameter $\xi$ are included by varying $\xi$ within its uncertainty and reevaluating the fits. These uncertainties are generally found to be negligible at the current level of statistical precision.

\section{Results}
\subsection{Energy splittings}

\begin{figure}
\centering
\includegraphics[width=0.48\linewidth]{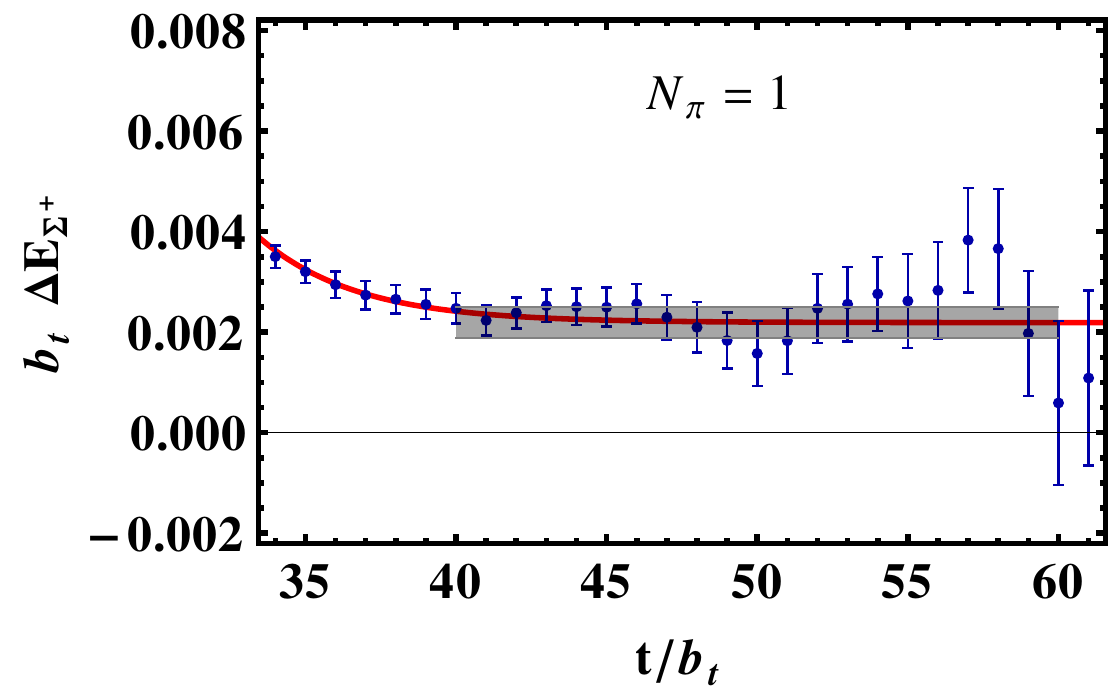}\hspace{1mm}
\includegraphics[width=0.48\linewidth]{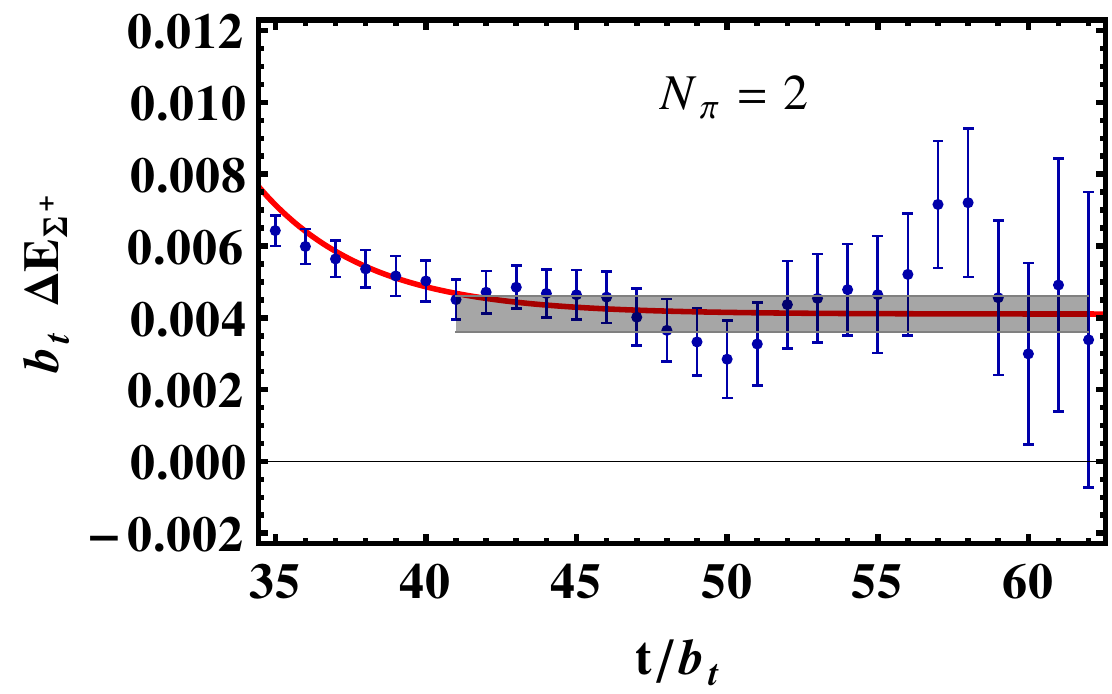}

\vspace{2mm}

\includegraphics[width=0.48\linewidth]{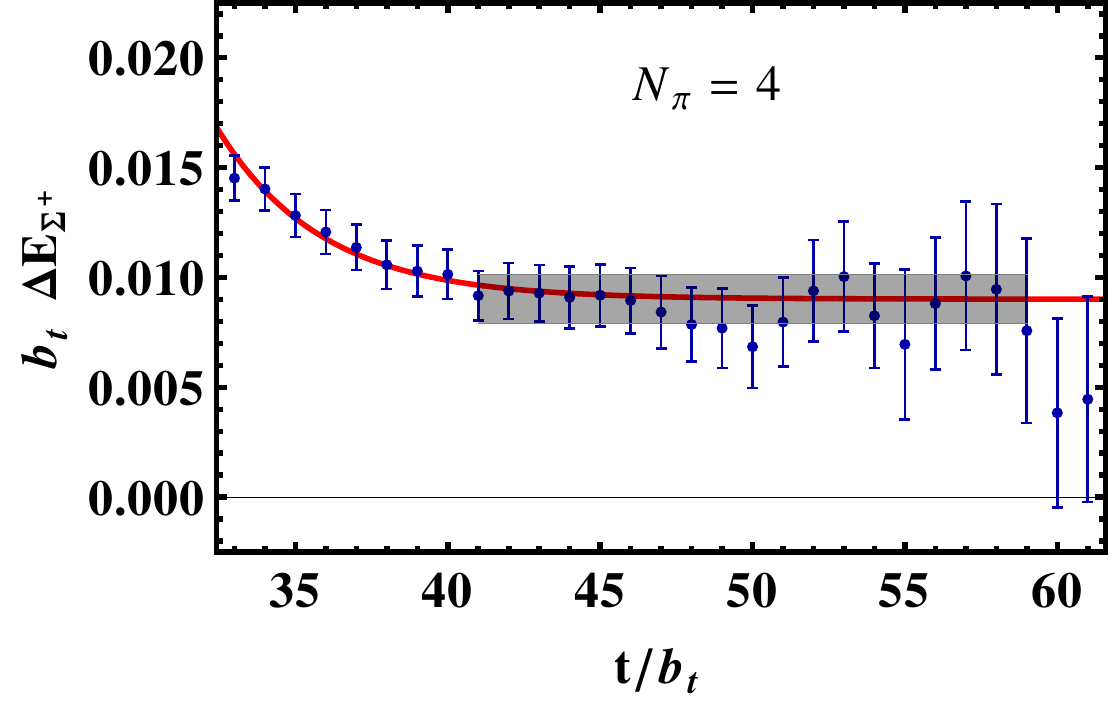}\hspace{1mm}
\includegraphics[width=0.48\linewidth]{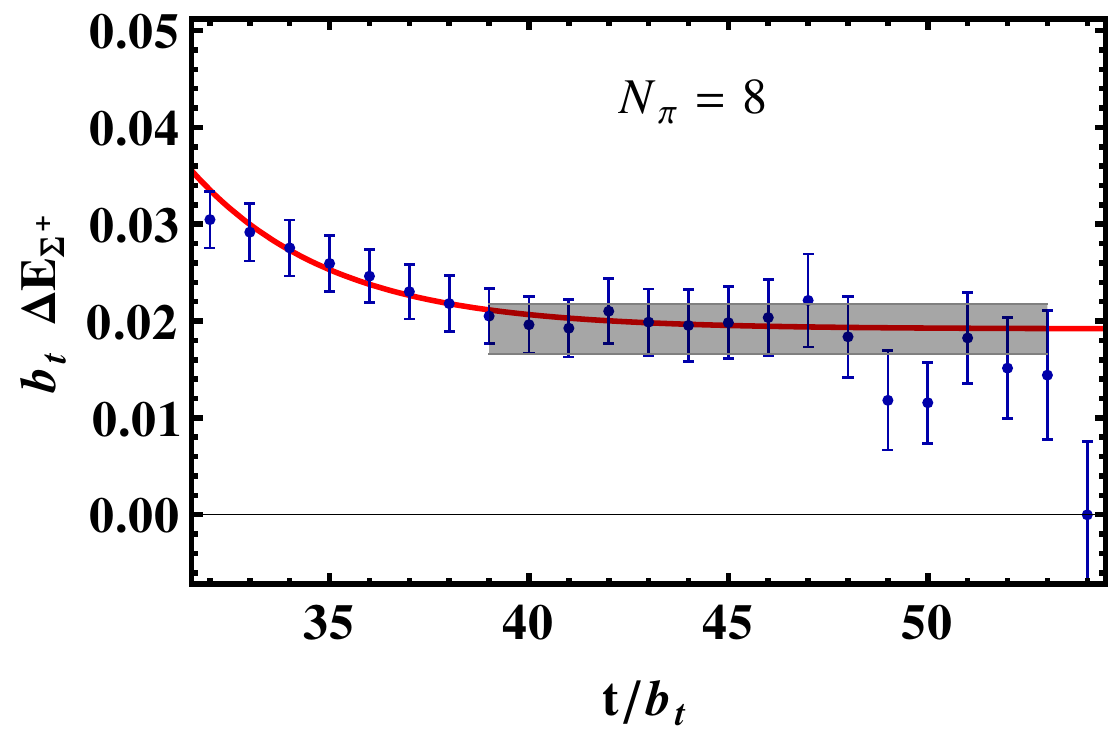}
\caption[]{\label{fig:res}Effective mass plots for the mass shift of the $\Sigma^{+}$ + $N_{\pi}$-pions systems. A fit to a constant plus exponential form is shown as a red line. The gray band represents the resulting energy extraction, with the fitting systematic uncertainty, obtained by varying the fitting endpoints by $\Delta \tau = \pm 2$, and the statistical uncertainty, combined in quadrature.}
\end{figure}

In \Figs{res}{reslast} we show sample effective mass plots for our baryon plus mesons systems, including those for the largest number of mesons for which we have a reliable signal for the energy shift. Included in the plots are results from a fit to a constant plus exponential, as well an error band showing the statistical and fitting systematic uncertainties combined in quadrature. From this data, we have extracted the mass splittings using the methods outlined above. The results from these extractions as a function of the number of mesons in the systems are shown in \Sec{ChiPT}, and numerical values are presented in \Tab{results}.

\begin{table}[]
\caption[]{\label{tab:results}Fitting results (in lattice units) for the energy shift of a single baryon in the presence of $n$ mesons. The first uncertainty is statistical, while the second represents a fitting systematic uncertainty obtained by changing the fitting endpoints by $\Delta \tau = \pm 2$.}
\begin{tabular}{|c|c|c|c|c|}
\hline
$n$ & $\Delta E\left(\Sigma^{+}(\pi^{+})^n\right)$ & $\Delta E\left(\Xi^{0}(\pi^{+})^n\right)$ & $\Delta E\left(p(K^{+})^n\right)$ & $\Delta E\left(n(K^{+})^n\right)$\\
\hline
1 &0.00219(22)(22) & 0.00057(19)(13) & 0.00242(22)(21) & 0.00068(25)(30)\\ 
2 &0.00410(46)(20) & 0.00162(45)(3) & 0.00481(40)(37) & 0.00166(46)(61) \\
3 &0.00670(56)(72) & 0.00273(70)(13) & 0.00726(57)(93) & 0.0029(9)(11) \\
4 &0.00901(80)(78) & 0.00367(84)(12) & 0.0103(9)(11) & 0.0038(13)(11) \\
5 &0.0113(11)(8) & 0.0050(12)(5) & 0.0131(11)(12) & 0.0048(12)(12) \\
6 &0.0137(14)(8) & 0.0067(15)(6) & 0.0165(14)(17) & - \\
7 &0.0169(17)(11) & 0.0080(21)(15) & 0.0198(17)(13) & - \\
8 &0.0192(14)(14) & - & 0.0224(19)(24) & - \\
9 &- & - & 0.0275(24)(20) & - \\
\hline
\end{tabular}
\end{table}

\begin{figure}
\centering
\includegraphics[width=0.48\linewidth]{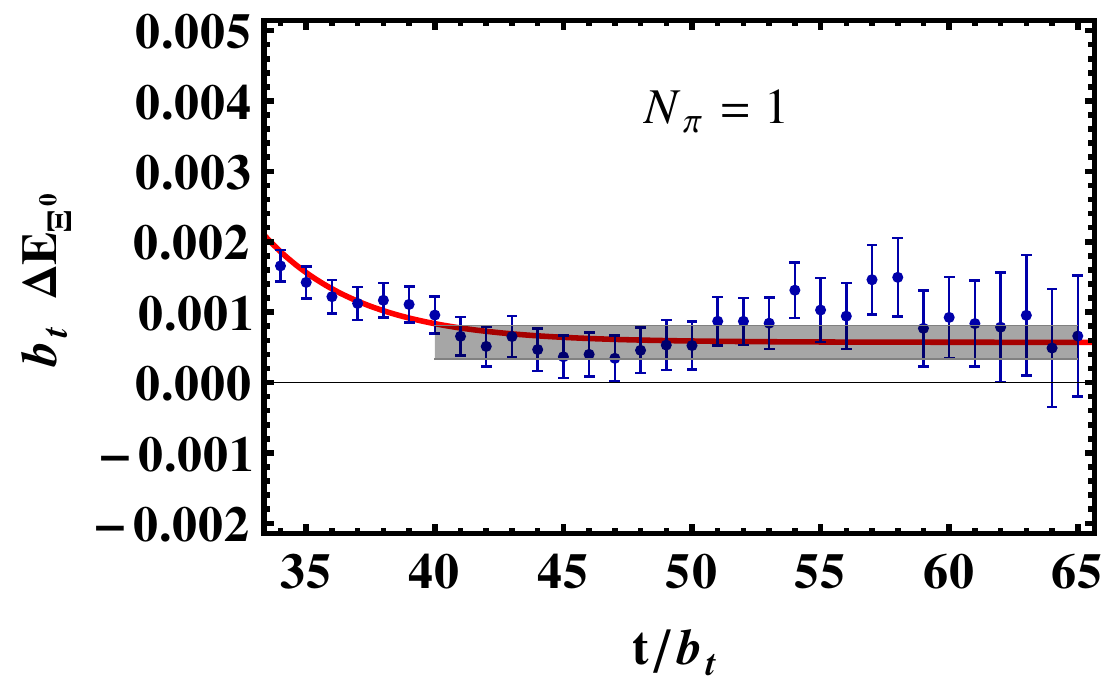}\hspace{1mm}
\includegraphics[width=0.48\linewidth]{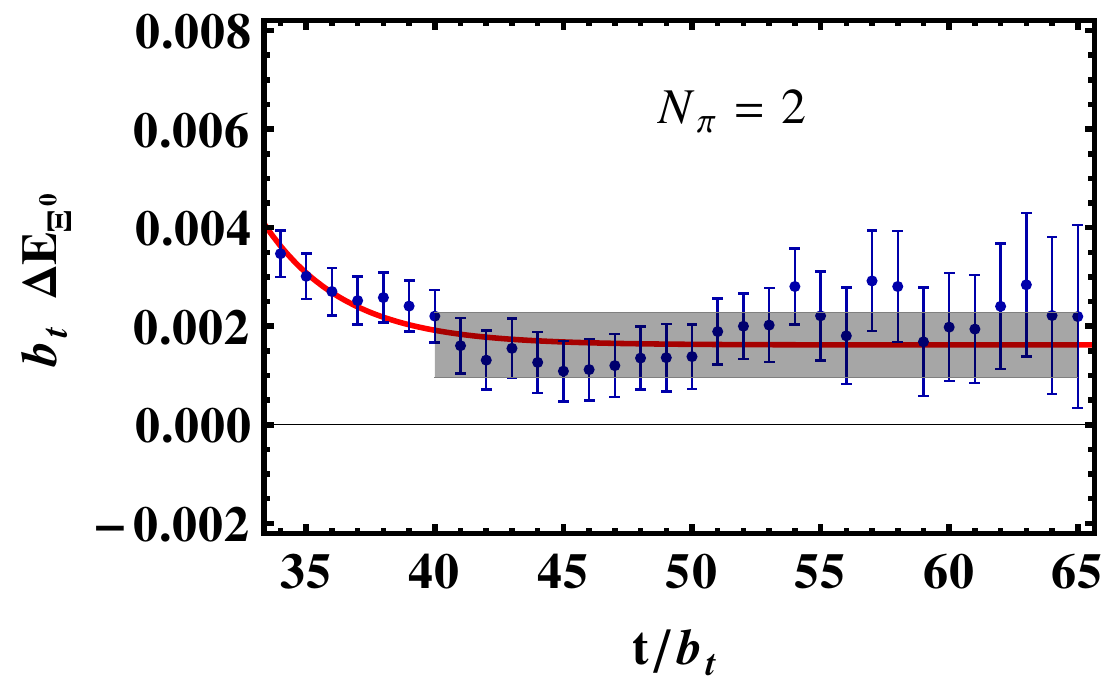}

\vspace{2mm}

\includegraphics[width=0.48\linewidth]{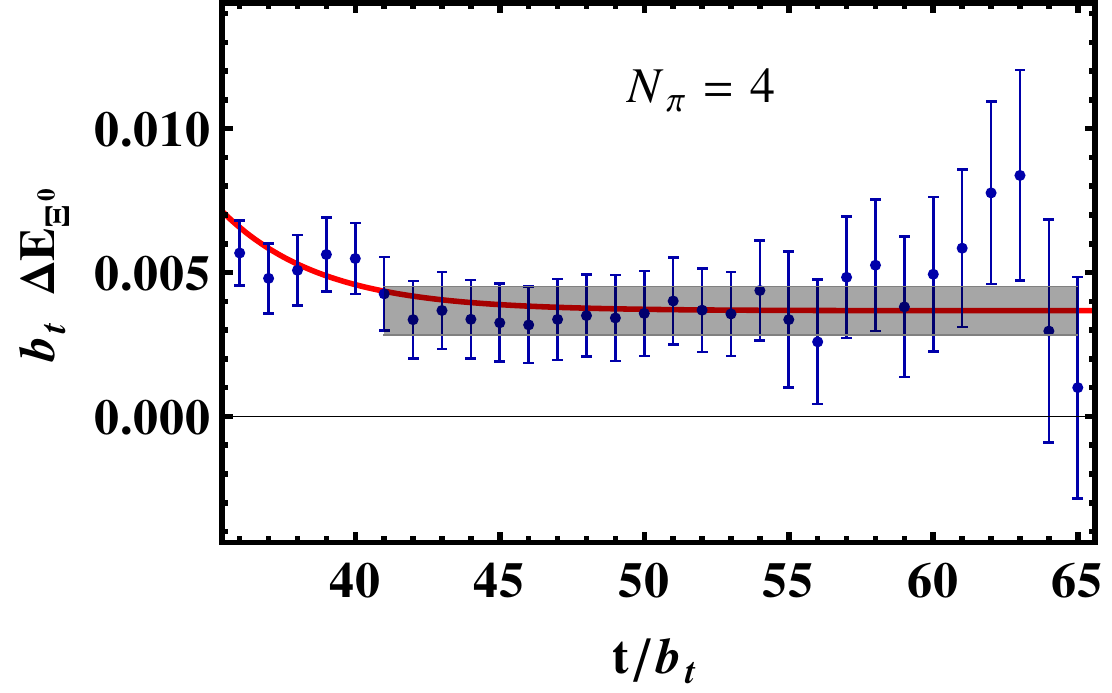}\hspace{1mm}
\includegraphics[width=0.48\linewidth]{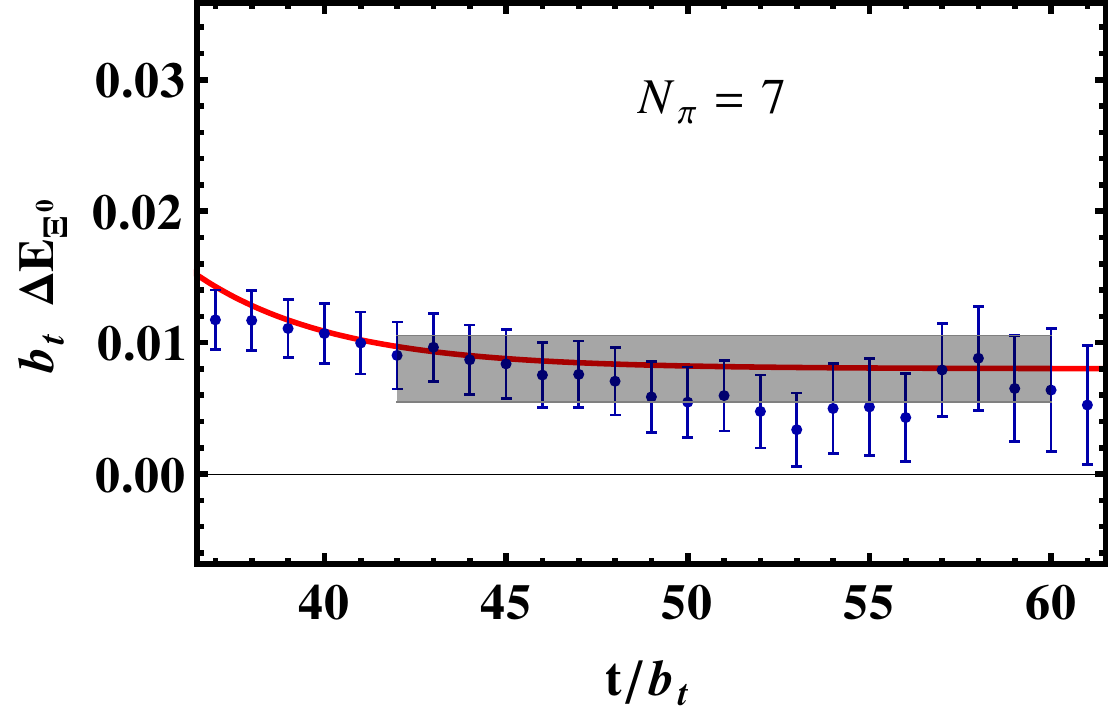}
\caption[]{\label{fig:reslast}Effective mass plots for the mass shift of the $\Xi^0$ + $N_{\pi}$-pions systems. A fit to a constant plus exponential form is shown as a red line. The gray band represents the resulting energy extraction, with the fitting systematic uncertainty, obtained by varying the fitting endpoints by $\Delta \tau = \pm 2$, and the statistical uncertainty, combined in quadrature.}
\end{figure}

\begin{figure}
\centering
\includegraphics[width=0.48\linewidth]{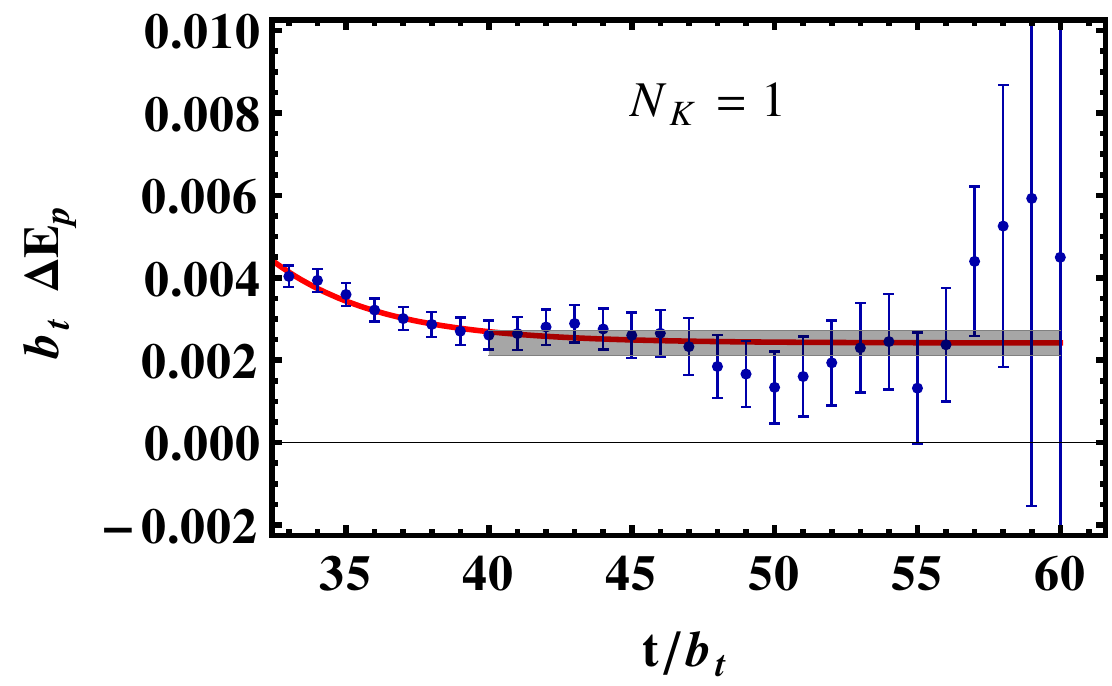}\hspace{1mm}
\includegraphics[width=0.48\linewidth]{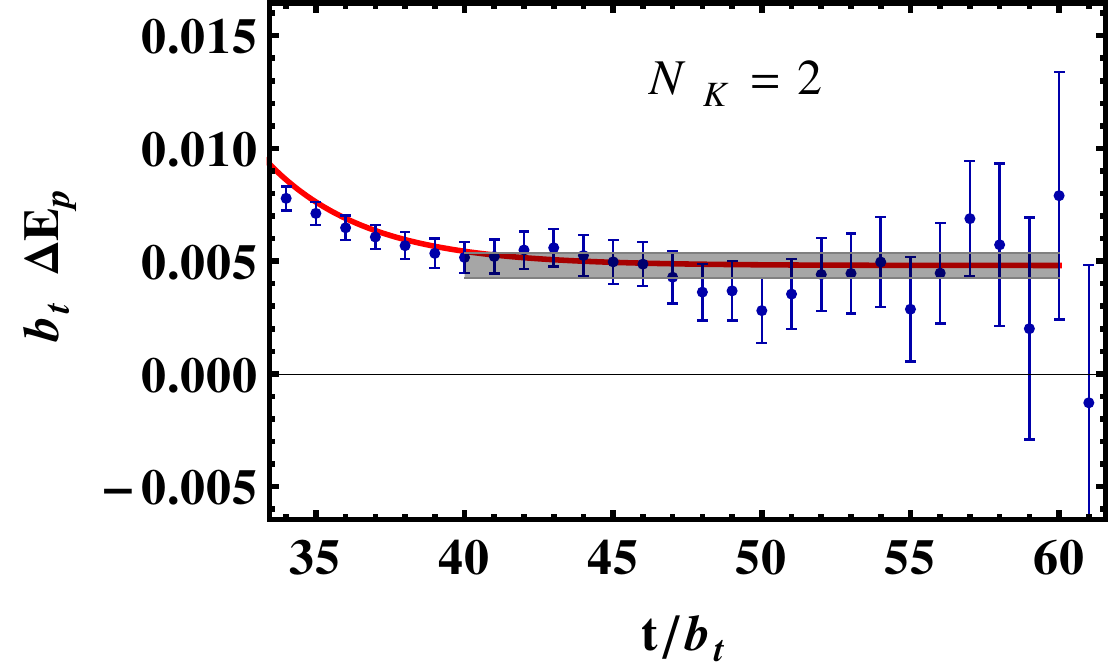}

\vspace{2mm}

\includegraphics[width=0.48\linewidth]{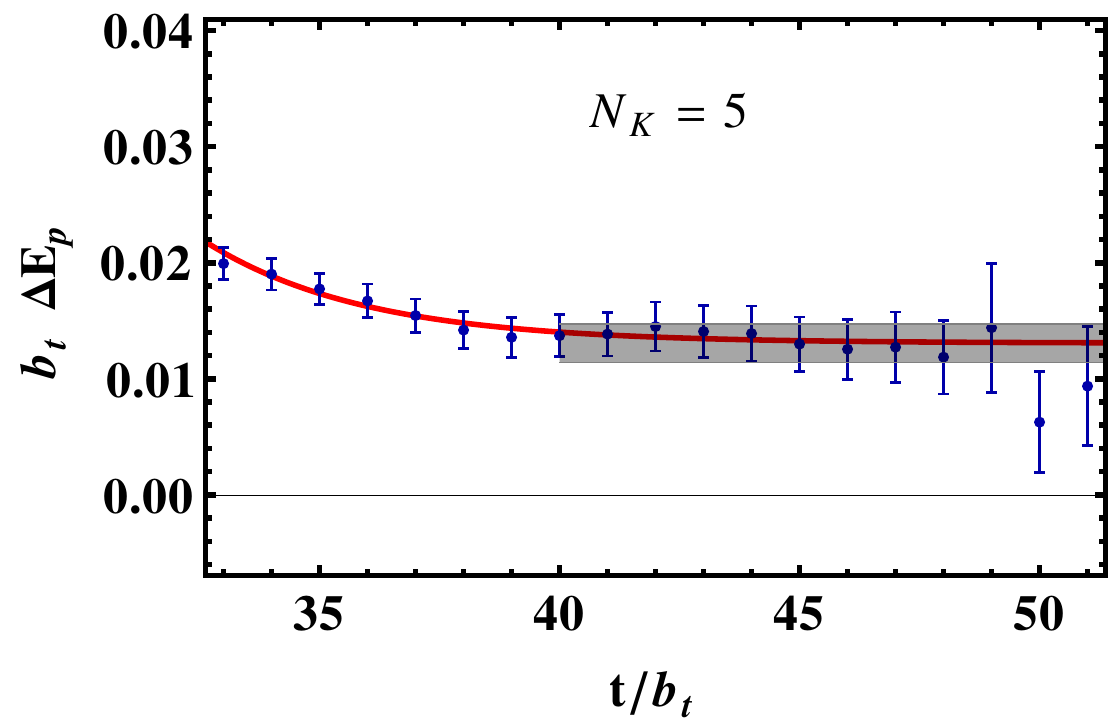}\hspace{1mm}
\includegraphics[width=0.48\linewidth]{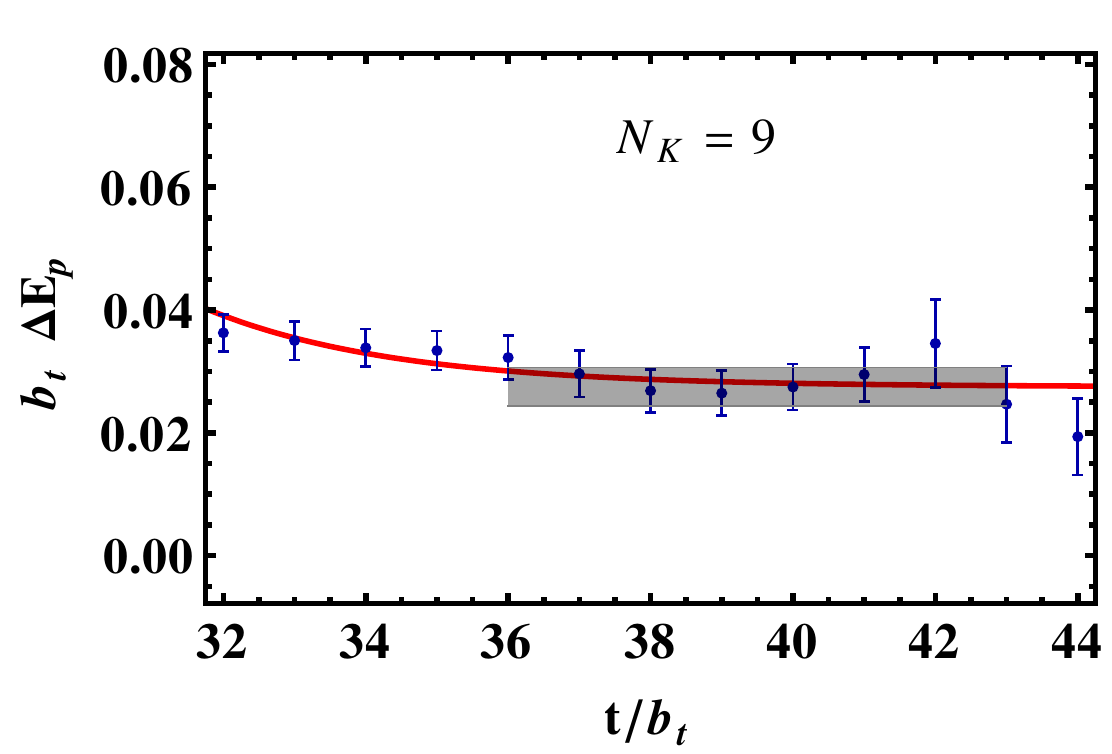}
\caption[]{Effective mass plots for the mass shift of the proton + $N_K$-kaons systems. A fit to a constant plus exponential form is shown as a red line. The gray band represents the resulting energy extraction, with the fitting systematic uncertainty, obtained by varying the fitting endpoints by $\Delta \tau = \pm 2$, and the statistical uncertainty, combined in quadrature.}
\end{figure}

\begin{figure}
\centering
\includegraphics[width=0.48\linewidth]{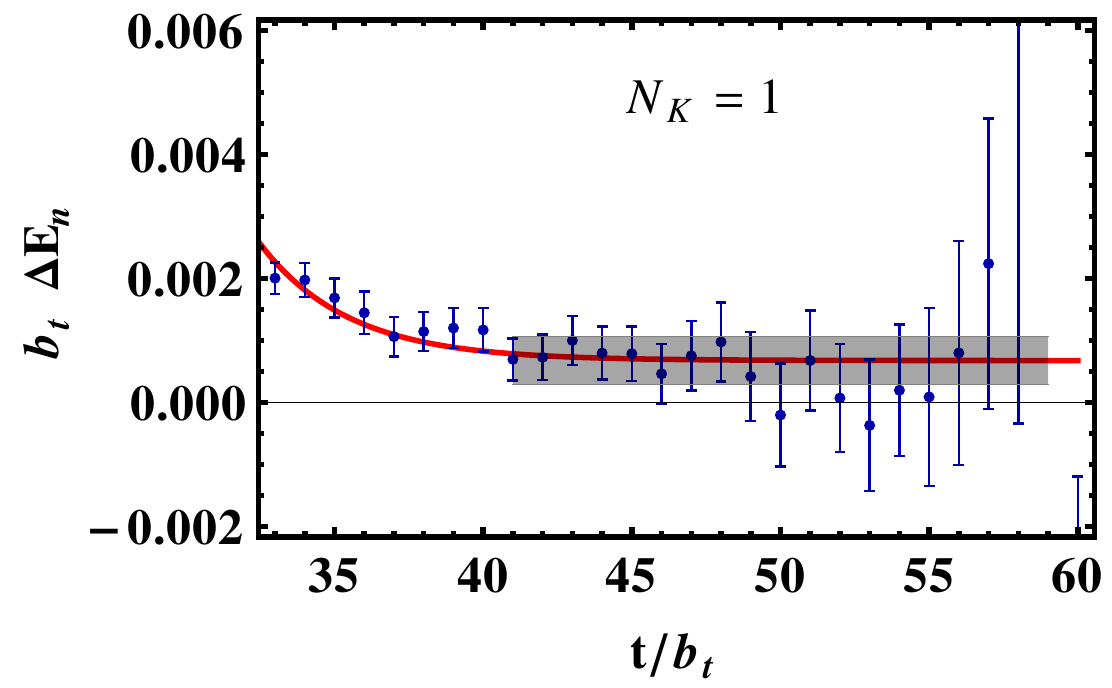}\hspace{1mm}
\includegraphics[width=0.48\linewidth]{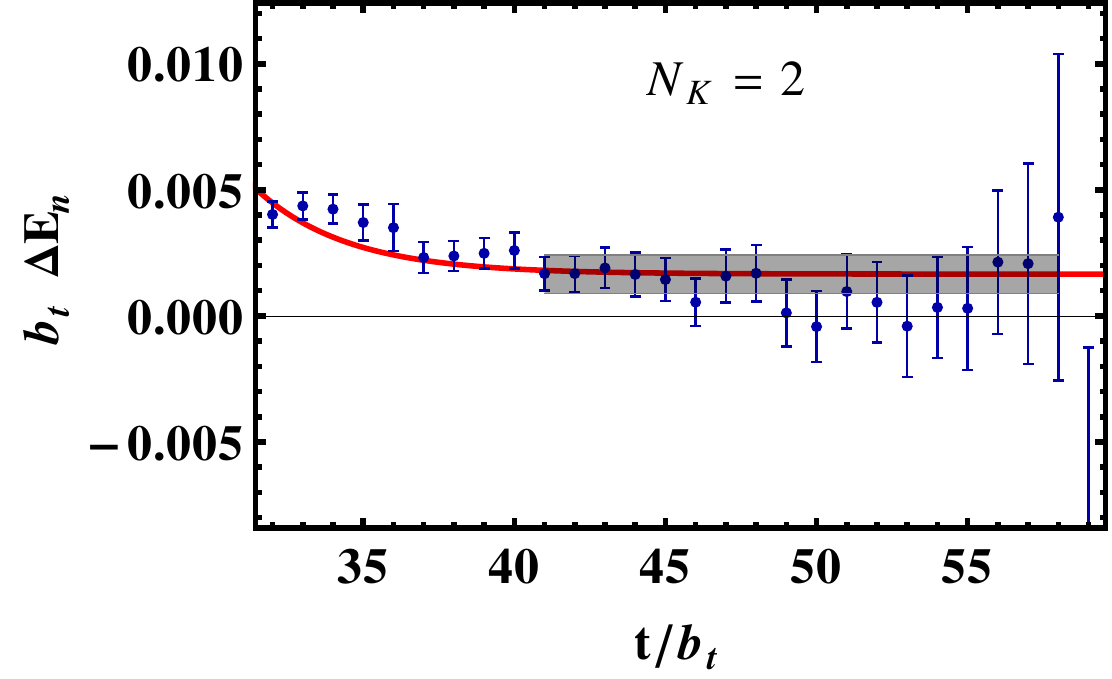}

\vspace{2mm}

\includegraphics[width=0.48\linewidth]{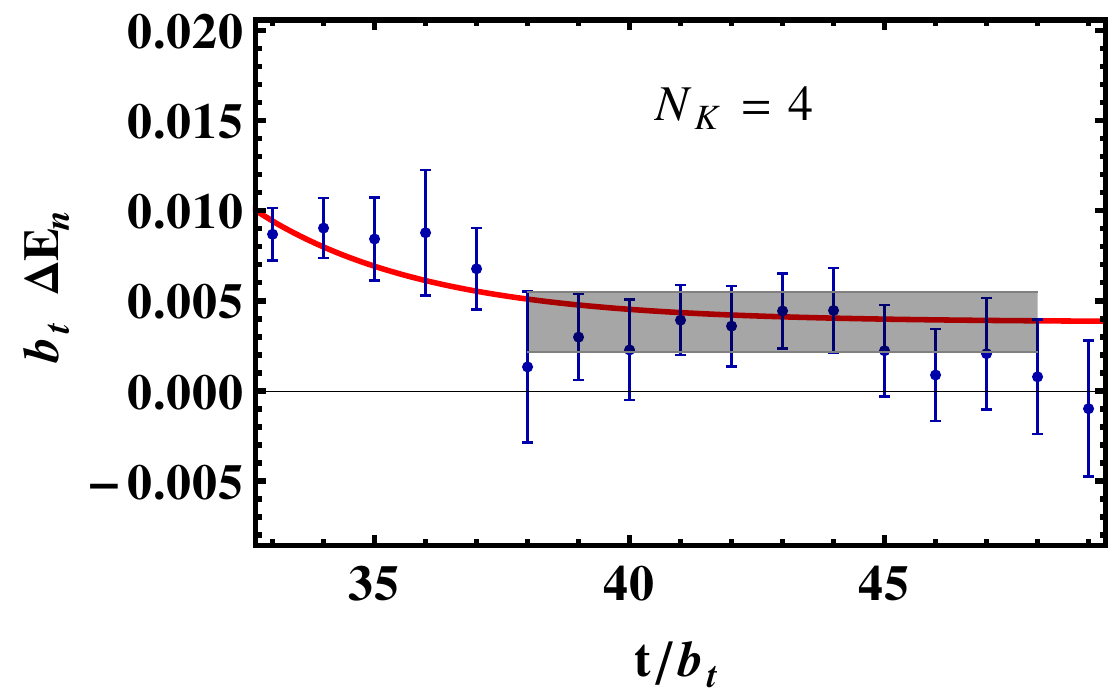}\hspace{1mm}
\includegraphics[width=0.48\linewidth]{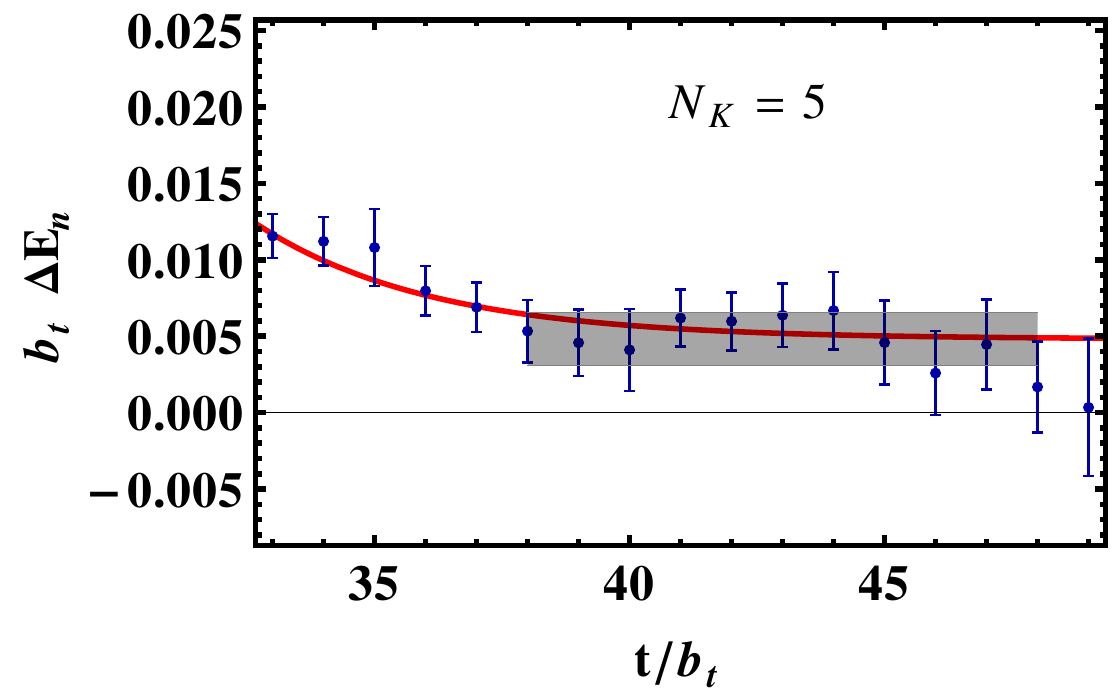}
\caption[]{\label{fig:reslast}Effective mass plots for the mass shift of the neutron + $N_K$-kaons systems. A fit to a constant plus exponential form is shown as a red line. The gray band represents the resulting energy extraction, with the fitting systematic uncertainty, obtained by varying the fitting endpoints by $\Delta \tau = \pm 2$, and the statistical uncertainty, combined in quadrature.}
\end{figure}

\subsection{Two- and three-body interactions}

Two- and three-body interactions have been extracted from the ground state energies using \Eq{LuscherExp}. Results for all two- and three-body interactions extracted from the lattice data are presented in \Tab{ScattPar}.

\begin{table}
\centering
\caption{\label{tab:ScattPar}Two- and three-body scattering parameters (in lattice units) extracted using \Eq{LuscherExp} for the meson-baryon systems and the result from \cite{Detmold:2008gh} for the meson-meson systems. For the central values we take the means of the results for each system size, $n$. Quoted uncertainties include statistical, fitting systematic uncertainties, as well as the standard deviation from the fits for all systems of $n$ mesons.}
\begin{tabular}{|c|c|c|c|c|}
\hline
 & $\bar{a}_{MM}$ & $\bar{a}_{MB}$ & $\bar{\eta}_{MMM}/10^{5}$ & $\bar{\eta}_{MMB}/10^{5}$ \\
\hline
$(\pi^{+})^n$ &1.191(70) &-&-1.2(1.1)&-\\
$\Sigma^{+} (\pi^{+})^n$ &1.194(92)&4.63(33)&-&3.4(3.4) \\
$\Xi^0 (\pi^{+})^n$ &1.189(94) &1.89(71)&-&8.0(2.7)\\
$(K^{+})^n$ & 1.565(79)&-&-0.21(48)&-\\
$p (K^{+})^n$ &1.562(83) &5.36(41)&-&6.2(3.1)\\
$n (K^{+})^n$ &1.552(84) &2.49(88)&-&8.6(4.5)\\
\hline
\end{tabular}
\end{table}

\begin{figure}
\centering
\includegraphics[width=0.48\linewidth]{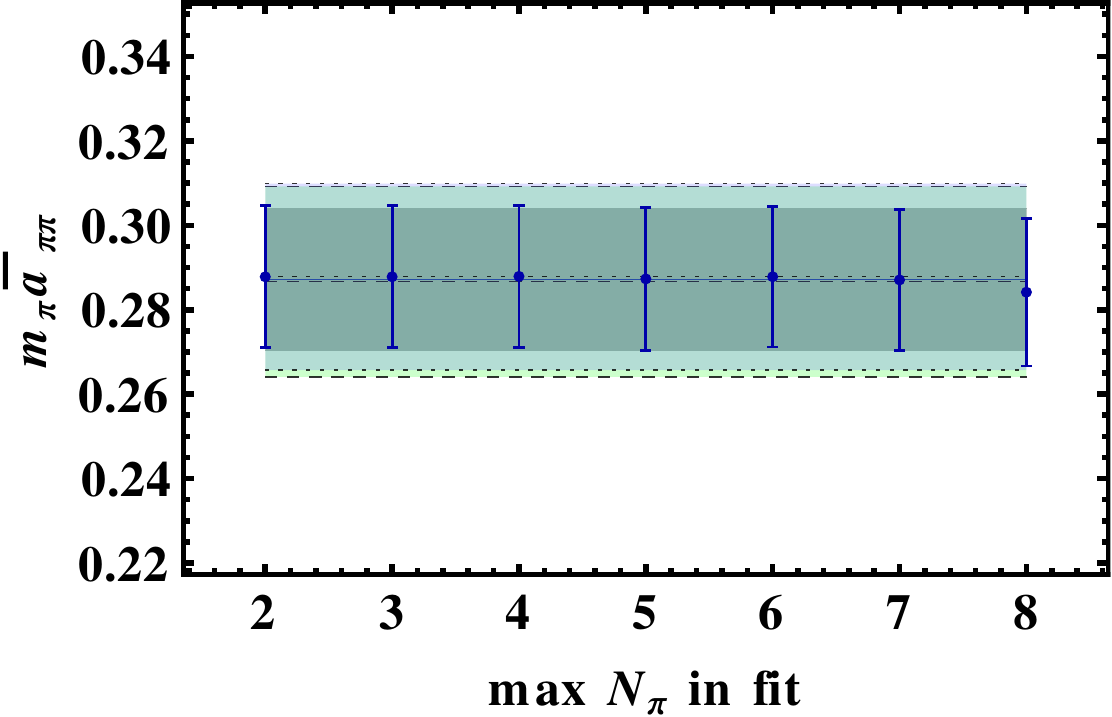}\hspace{2mm}
\includegraphics[width=0.48\linewidth]{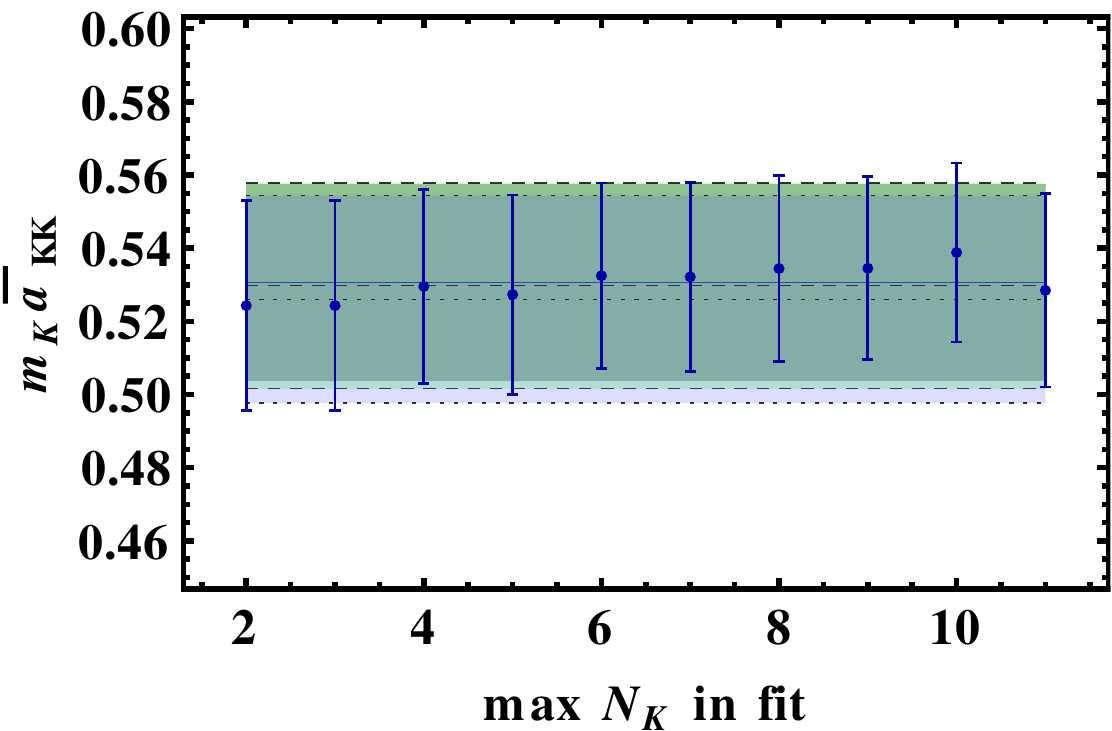}
\caption{\label{fig:aPiPi}Inverse scattering phase shifts, $\bar{a} = \left(p\cot\delta\right)^{-1}$ for pions (left) and kaons (right), as a function of the maximum system size included in the fit. Error bars represent the statistical and systematic uncertainties from the individual fits combined in quadrature. The gray band shows the mean of all fits and their uncertainties, plus an additional uncertainty given by the standard deviation of all fits, added in quadrature. The blue and green bands (outlined by dotted and dashed lines, respectively) show the results from the extraction of the same parameters from the $\pi^{+}\Sigma^{+}$ and $\pi^{+}\Xi^0$ systems, respectively (left), and from the $K^{+}$n and $K^{+}$p systems, respectively (right).}
\end{figure}

We may extract the $I=2$ $\pi\pi$ and $I=1$ $KK$ two-body interactions using the mixed species formula presented above for the meson baryon systems, as well as from the single species formula \cite{Detmold:2008gh}, using the data from $C_{n}(t)$, which were also computed in order to calculate the energy splittings for the meson baryon systems. The results for the two- and three-meson interactions are shown as a function of the maximum number of mesons included in the fit in \Figtwo{aPiPi}{etaPiK}. We see that the extractions from the mixed and pure ensembles are consistent, and that the parameters remain consistent as the system size is varied. 

\begin{figure}
\centering
\includegraphics[width=0.48\linewidth]{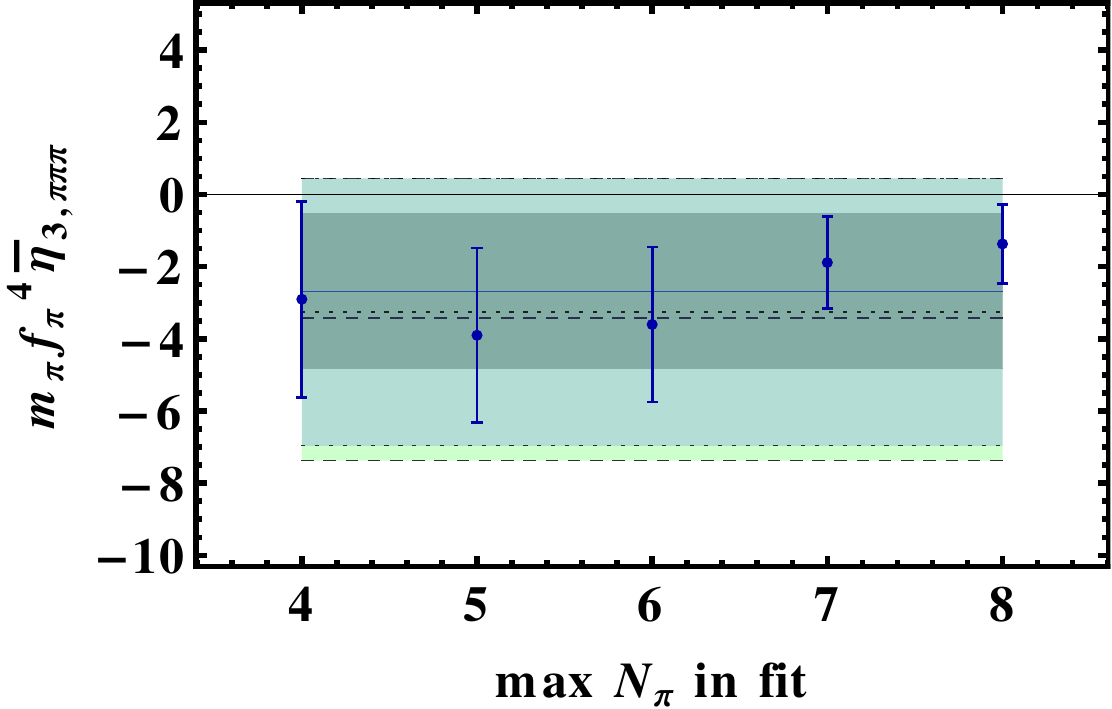}\hspace{1mm}
\includegraphics[width=0.48\linewidth]{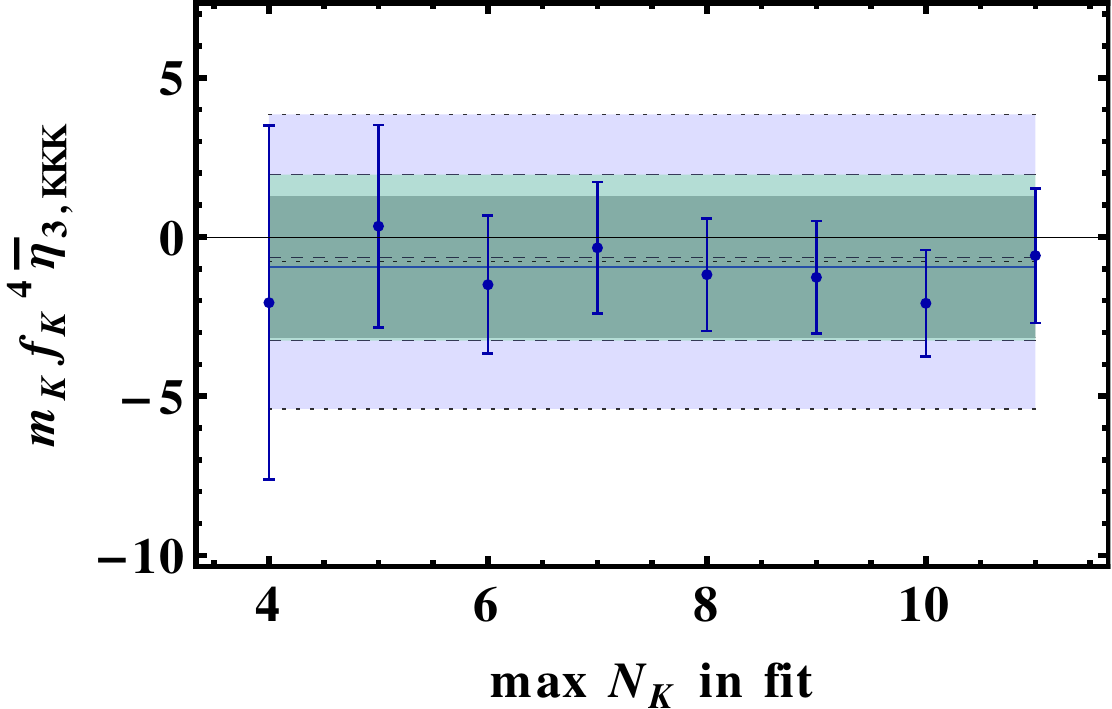}
\caption[.]{\label{fig:etaPiK}Three-body interaction parameters, $\bar{\eta}_{3,\pi\pi\pi}$ (left) and $\bar{\eta}_{3,KKK}$ (right), as a function of the maximum system size included in the fit. Error bars represent the statistical and systematic uncertainties from the individual fits combined in quadrature. The gray band shows the mean of all fits and their uncertainties, plus an additional uncertainty given by the standard deviation of all fits, added in quadrature. The blue and green bands (outlined by dotted and dashed lines, respectively) show the results from the extraction of the same parameters from the $\pi^{+}\Sigma^{+}$ and $\pi^{+}\Xi^0$ systems, respectively (left), and from the $K^{+}$n and $K^{+}$p systems, respectively (right).}
\end{figure}

\begin{figure}
\centering
\includegraphics[width=0.49\linewidth]{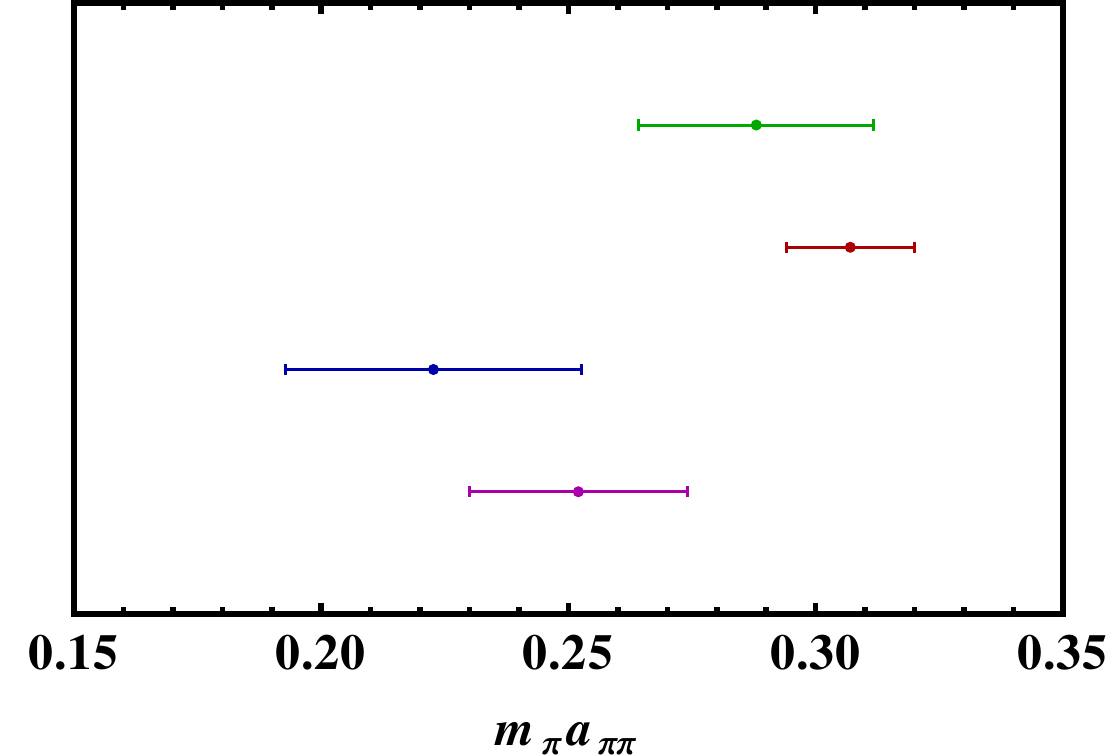}\hspace{1mm}
\includegraphics[width=0.49\linewidth]{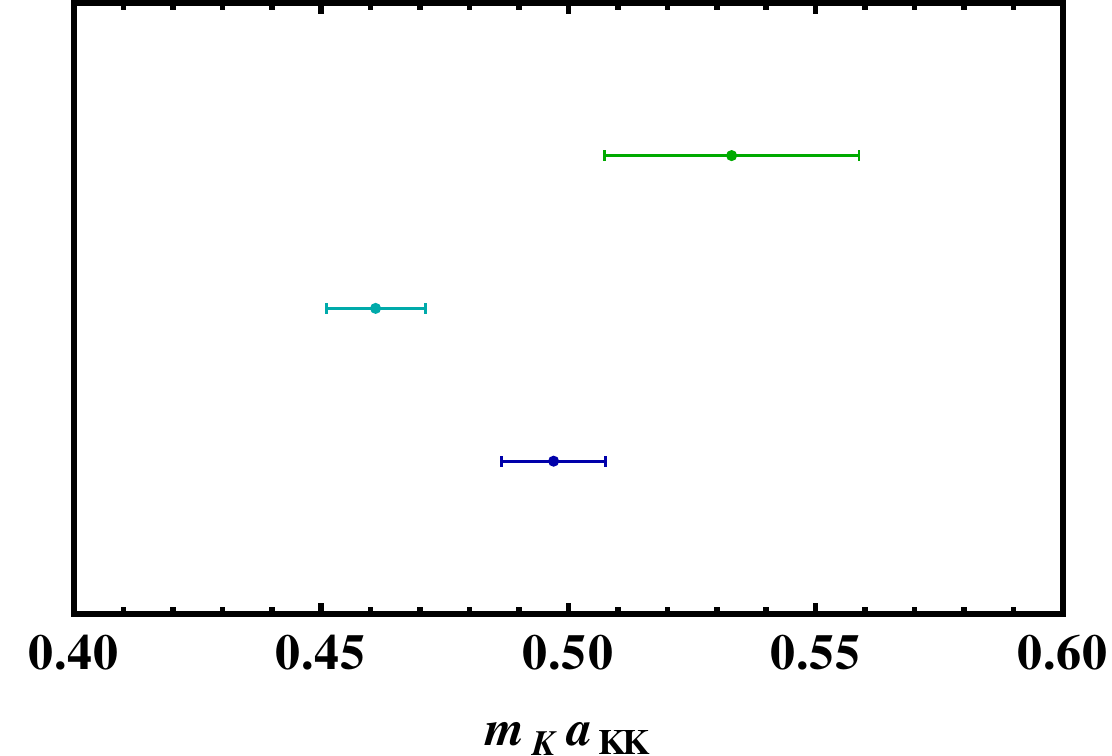}
\caption{\label{fig:mpiapipi}Scattering lengths for $I=2$ $\pi\pi$ (left) and $I=1$ $KK$ (right) extracted by various groups with pion masses near 350 MeV. From bottom to top (left): ETMC \cite{Feng:2009ij}, NPLQCD \cite{Beane:2011sc}, HSC \cite{Dudek:2012gj}, this work. From bottom to top (right): NPLQCD \cite{Detmold:2008yn}, Detmold \& Smigielski \cite{Detmold:2011kw}, this work. In \cite{Beane:2011sc,Dudek:2012gj}, the $\pi\pi$ scattering length is extracted from an extrapolation to threshold; all other points represent effective scattering lengths, $\bar{a}$.}
\end{figure}

In \Fig{mpiapipi}, we compare our results for the effective scattering lengths to calculations performed by other groups at similar pion masses. Also included in this plot are two calculations in which the $\pi\pi$ scattering length is extracted from an extrapolation to threshold. We find no significant discrepancies between this work and that of previous efforts within the quoted uncertainties. Finally, we find a nonzero three-pion interaction and no three-kaon interaction within our current uncertainty. Previous work by the NPLQCD collaboration also found nonzero three-pion interactions (\cite{Beane:2007es,Detmold:2008fn,Detmold:2011kw,Detmold:2012wc}), however, the sign of the three-pion interaction in our case is opposite to that found for smaller volumes in these previous works. The volume-dependence of the three-pion interaction has thus far not been found to be consistent with the perturbative prediction \cite{Detmold:2012wc}, and warrants further investigation to understand its behavior.

\begin{figure}
\centering
\includegraphics[width=0.48\linewidth]{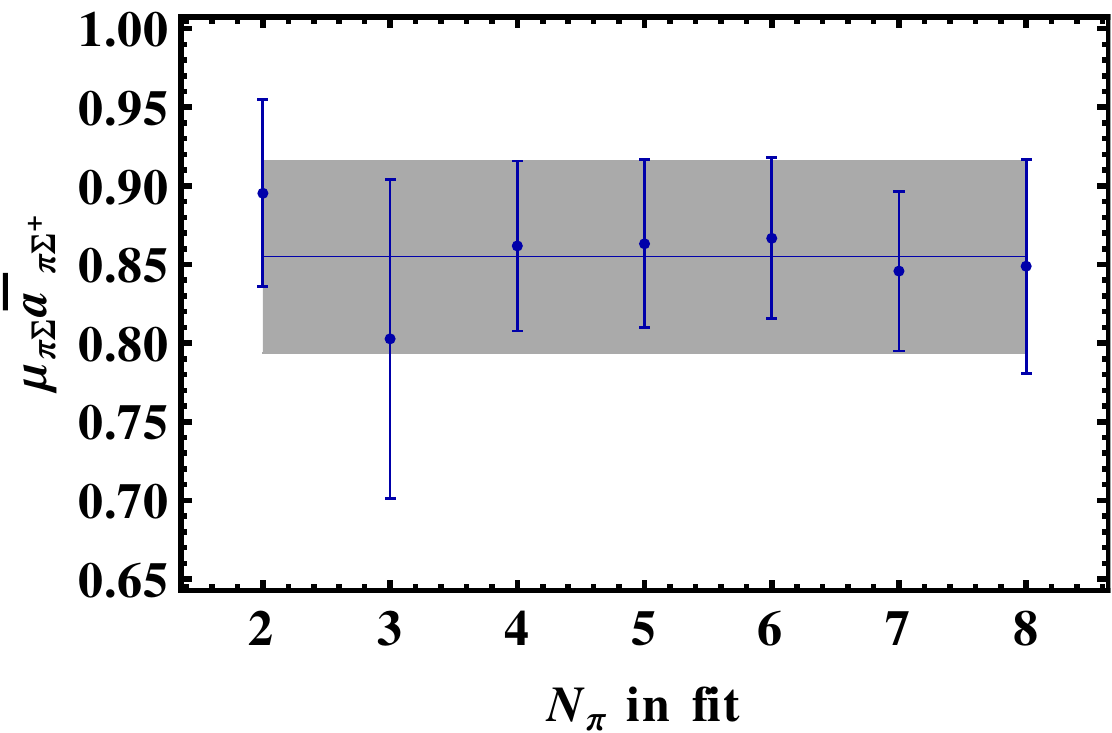}\hspace{1mm}
\includegraphics[width=0.48\linewidth]{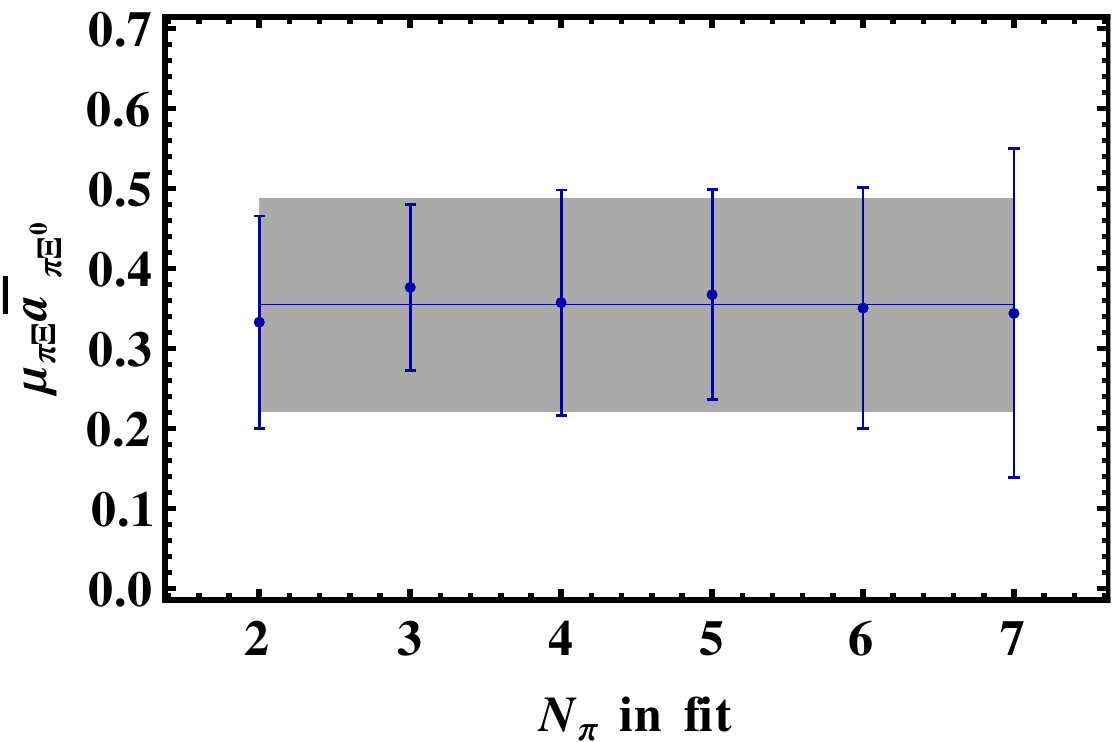}\\

\vspace{2mm}

\includegraphics[width=0.48\linewidth]{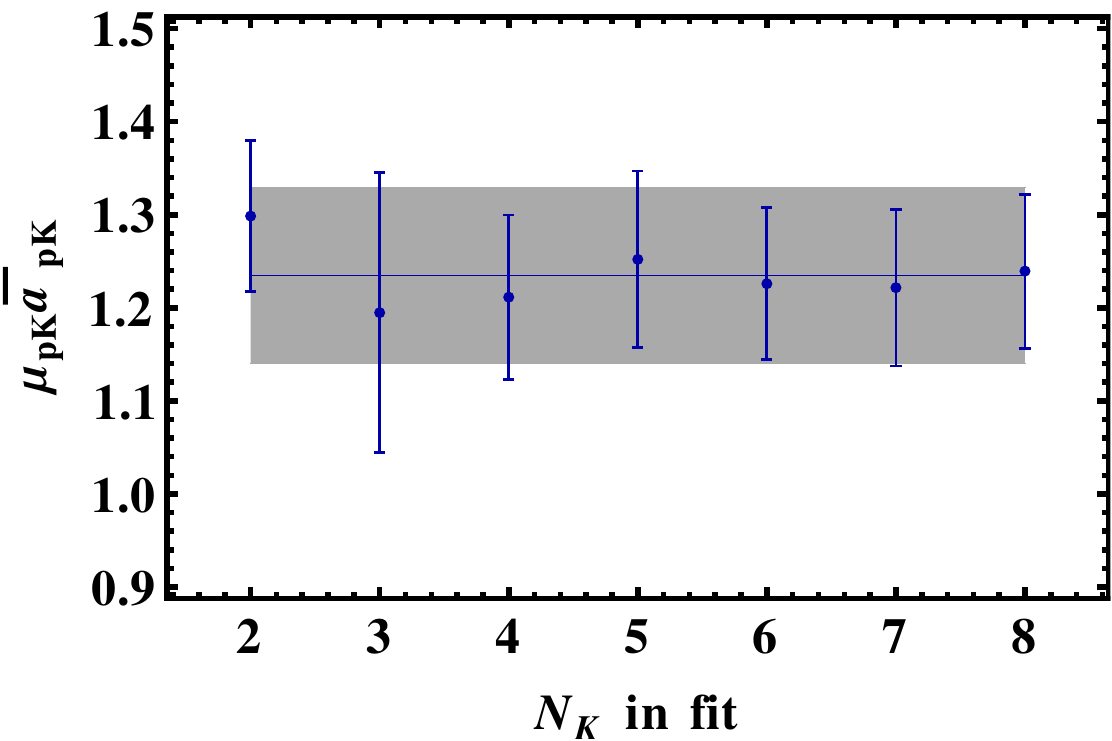}\hspace{1mm}
\includegraphics[width=0.48\linewidth]{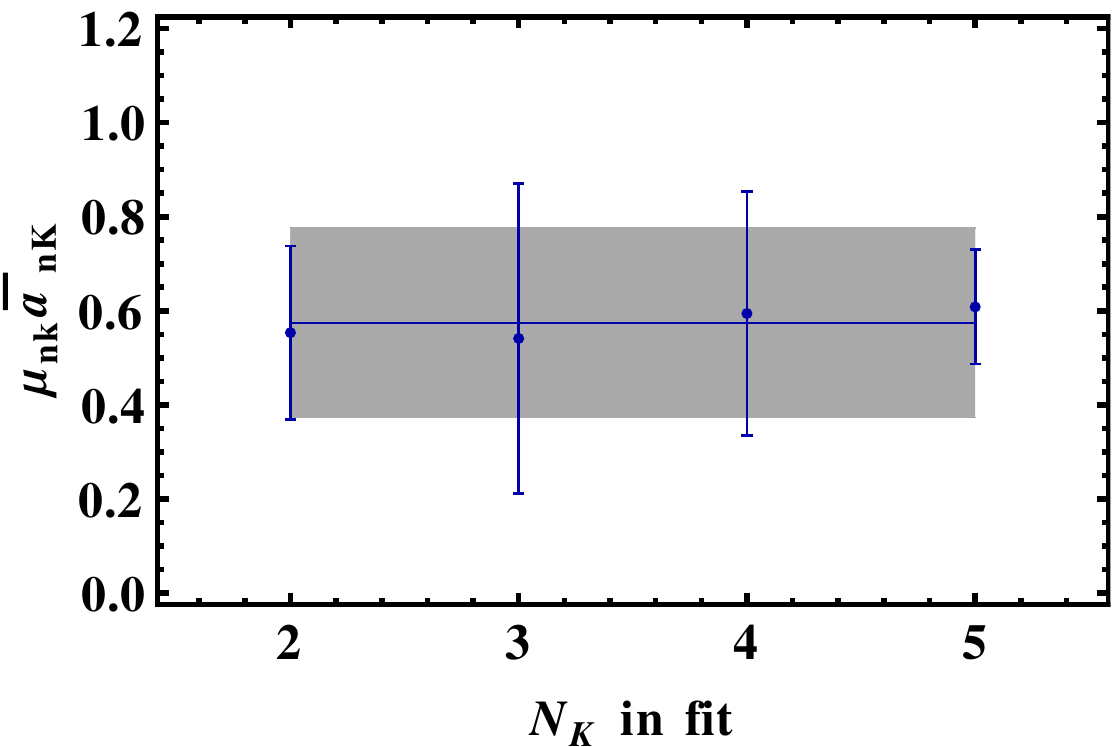}
\caption{\label{fig:aMB}Inverse scattering phase shifts, $\bar{a} = \left(p\cot\delta\right)^{-1}$, as a function of the maximum system size included in the fit. Clockwise from upper left: $\pi^{+}\Sigma^{+}$, $\pi^{+}\Xi^0$, $K^{+}$n, and $K^{+}$p. Error bars represent the statistical and systematic uncertainties from the individual fits combined in quadrature. The gray band shows the mean of all fits and their uncertainties, plus an additional uncertainty given by the standard deviation of all fits, added in quadrature.}
\end{figure}

\begin{figure}
\centering
\includegraphics[width=0.48\linewidth]{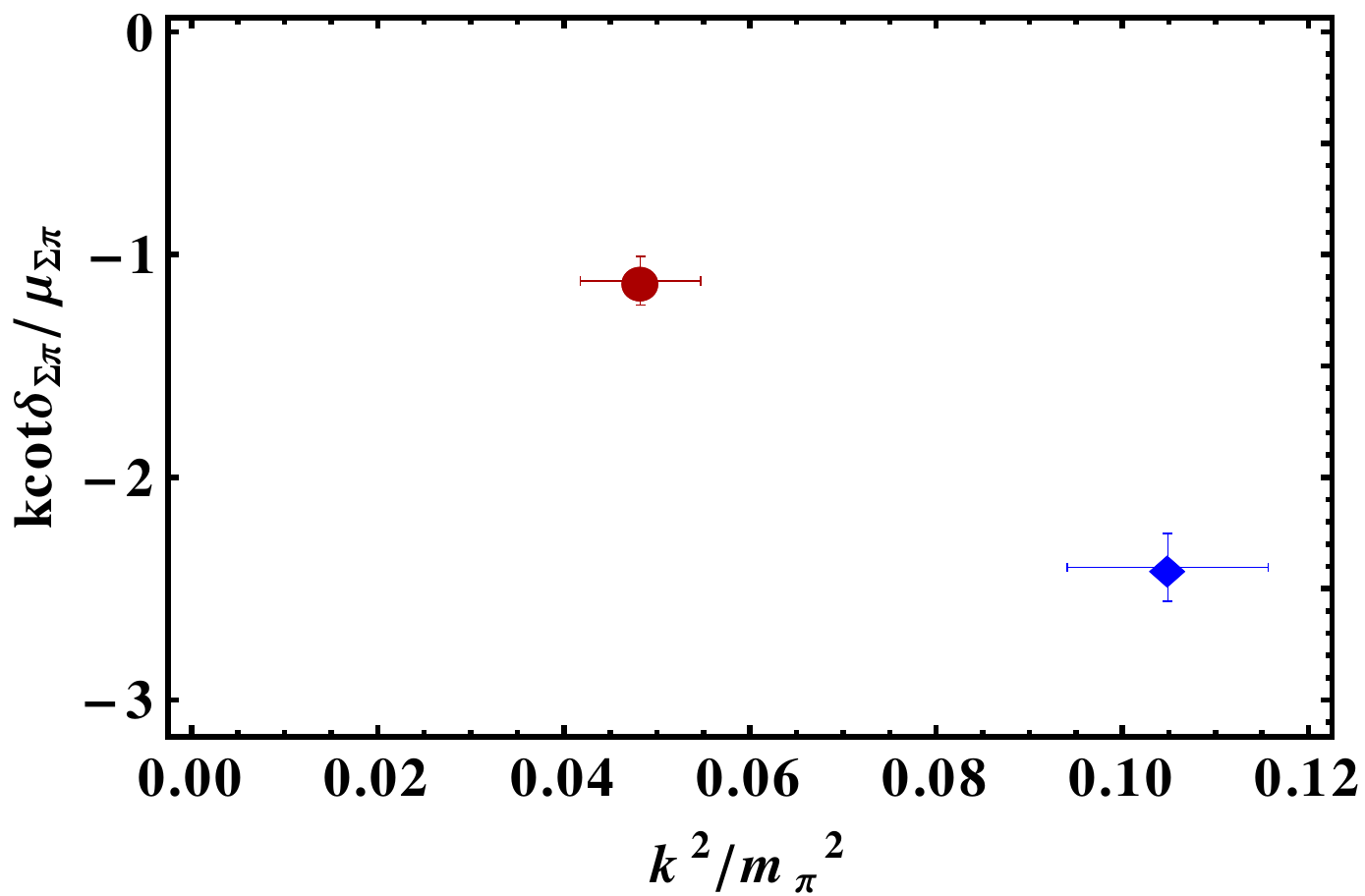} \hspace{1mm}
\includegraphics[width=0.48\linewidth]{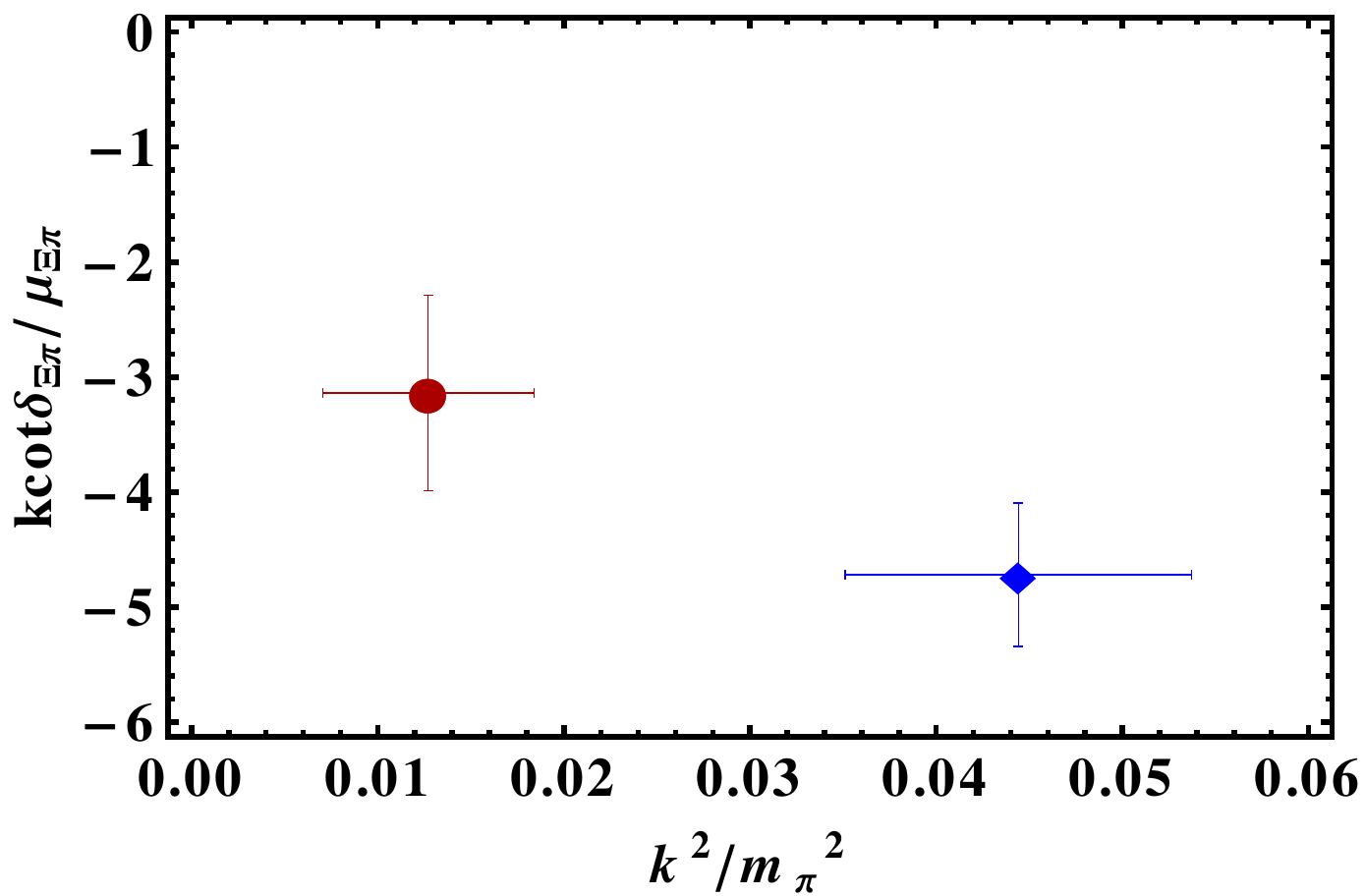} \\

\vspace{2mm}

\includegraphics[width=0.48\linewidth]{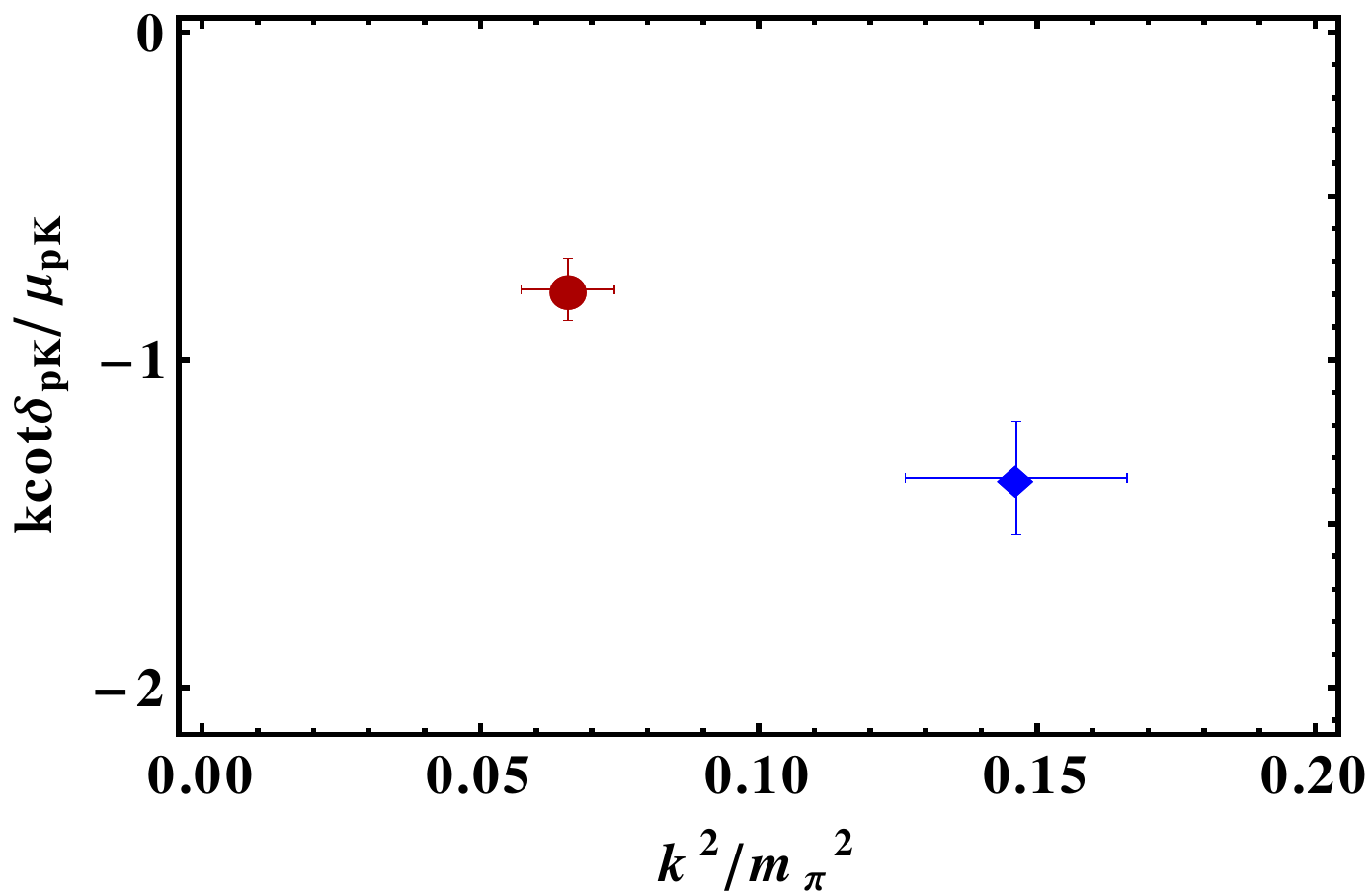}\hspace{1mm}
\includegraphics[width=0.48\linewidth]{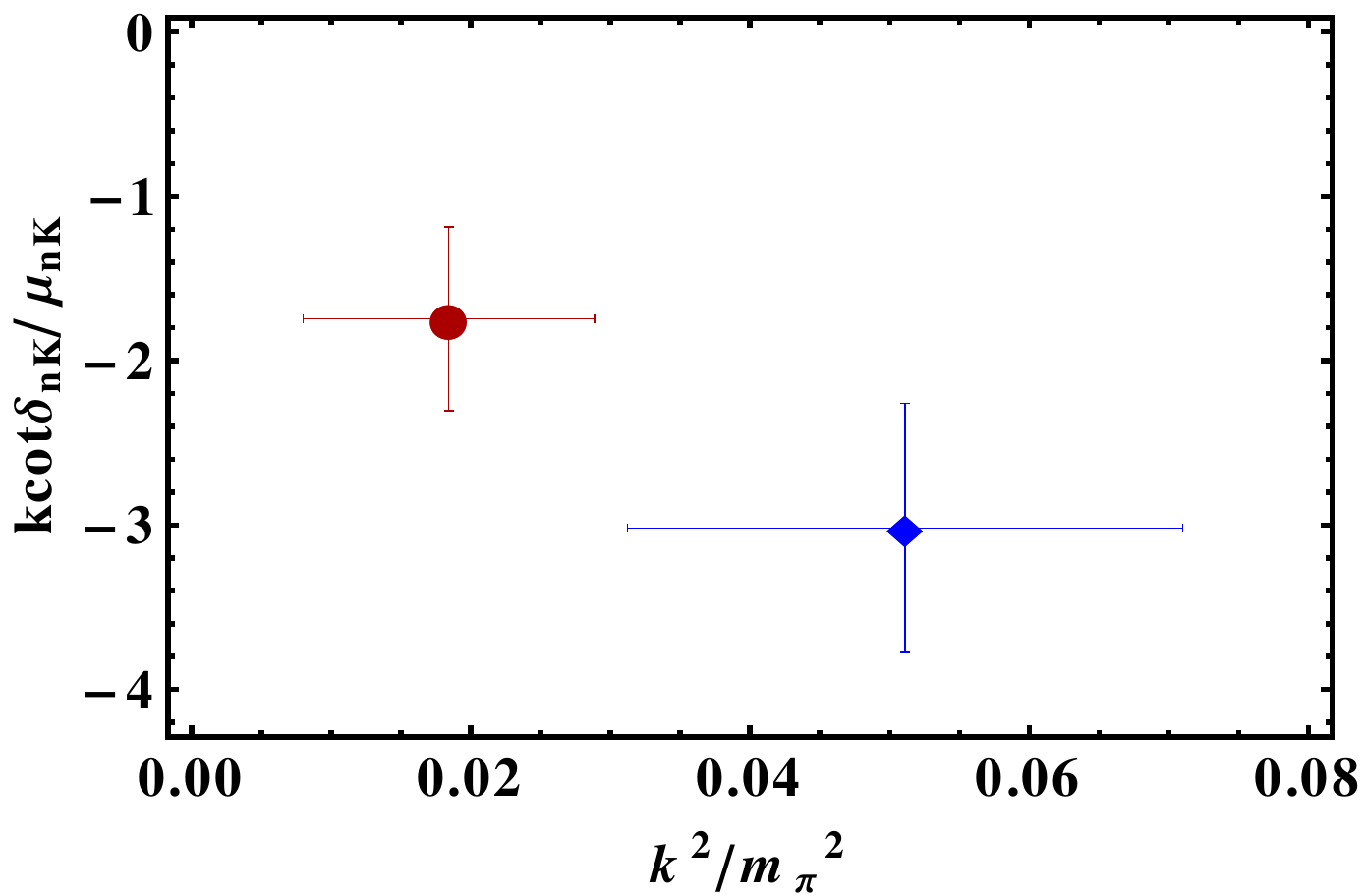}
\caption{\label{fig:pcotd}Scattering phase shifts, $k\cot\delta$, vs. scattering momenta, $k^2$, calculated in this work (red circles) and by NPLQCD \cite{Torok:2009dg} (blue diamonds). Clockwise from upper left: $\pi^{+}\Sigma^{+}$, $\pi^{+}\Xi^0$, $K^{+}$n, and $K^{+}$p.}
\end{figure}

Our extracted two-body meson-baryon interactions are shown in \Fig{aMB} as a function of the number of mesons included in the fit. Again, we find no variation with system size. In \Fig{pcotd}, we plot our extracted $p\cot\delta$ as a function of the scattering momentum $p$, along with the results from Ref.~\cite{Torok:2009dg}, performed at a smaller volume, and therefore a correspondingly larger scattering momentum. We find that there is still significant variation of the scattering phase shift even at rather small momenta compared to the pion mass. This signifies that contributions from short-distance physics, such as the effective range, $t$-channel cuts, or inelasticities, are relevant at these momenta. Extraction of the effective range requires studies at additional volumes and will be the subject of future work.

\begin{figure}
\centering
\includegraphics[width=0.48\linewidth]{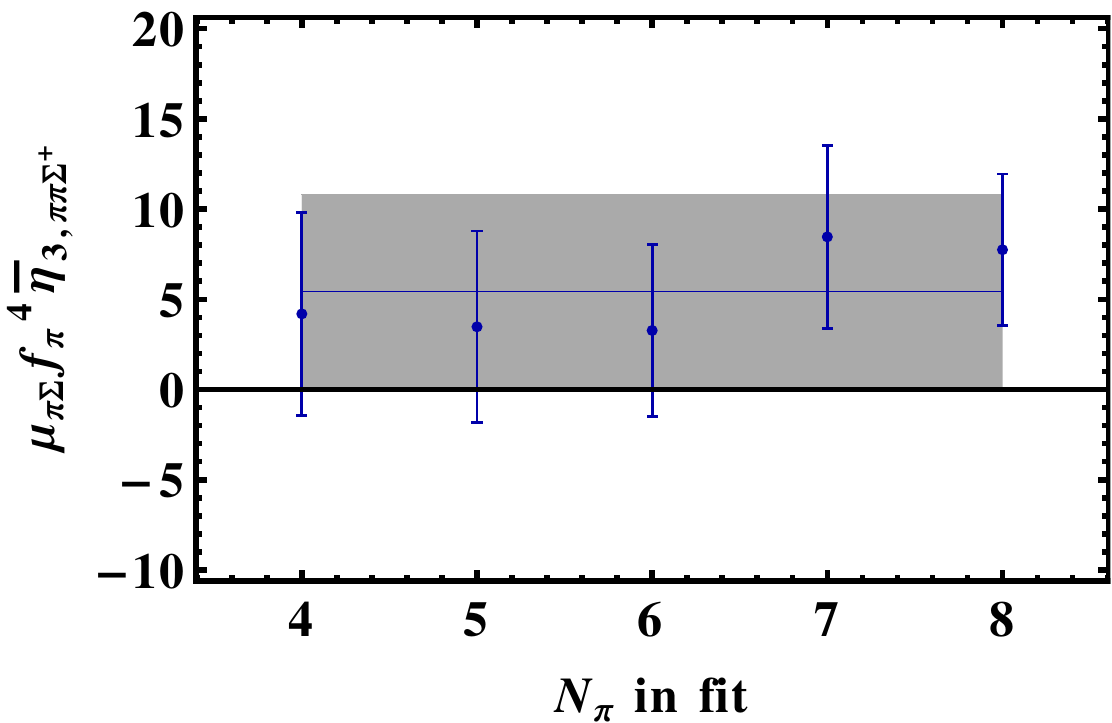}\hspace{1mm}
\includegraphics[width=0.48\linewidth]{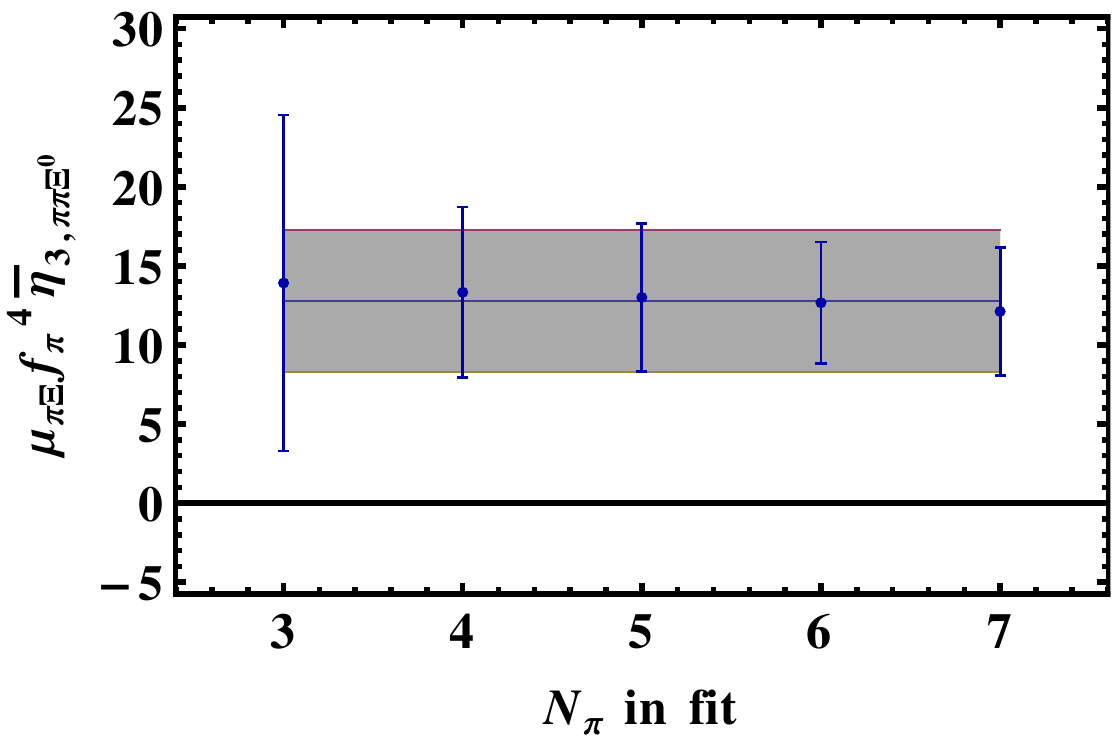} \\

\vspace{2mm}

\includegraphics[width=0.48\linewidth]{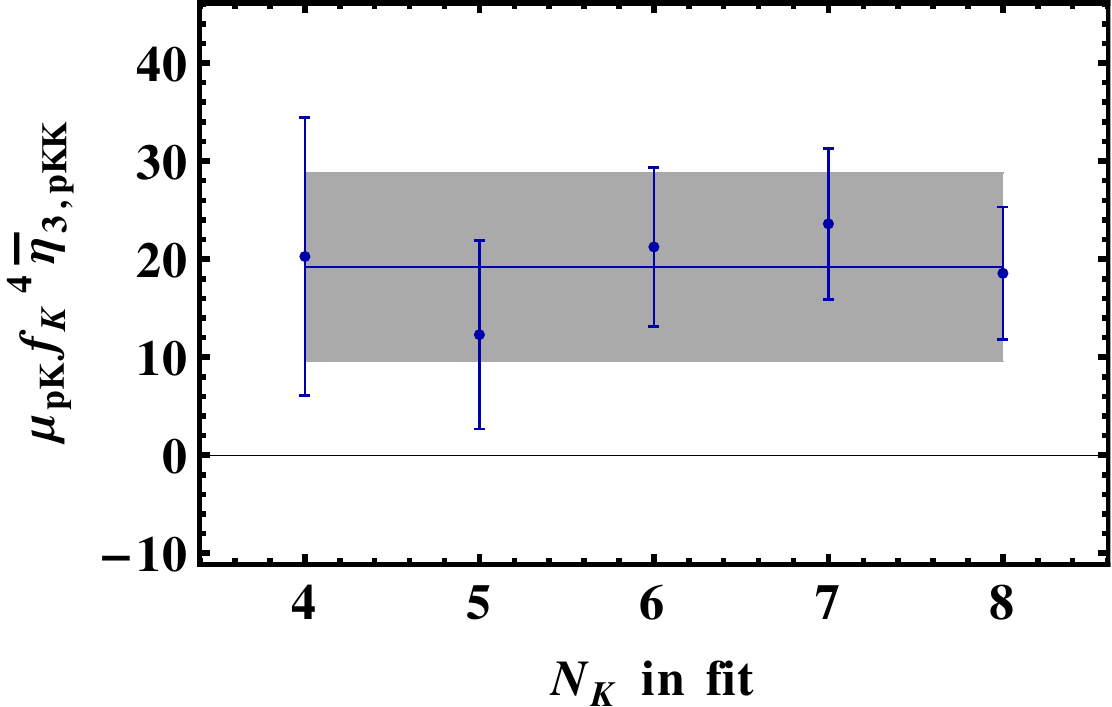}\hspace{1mm}
\includegraphics[width=0.48\linewidth]{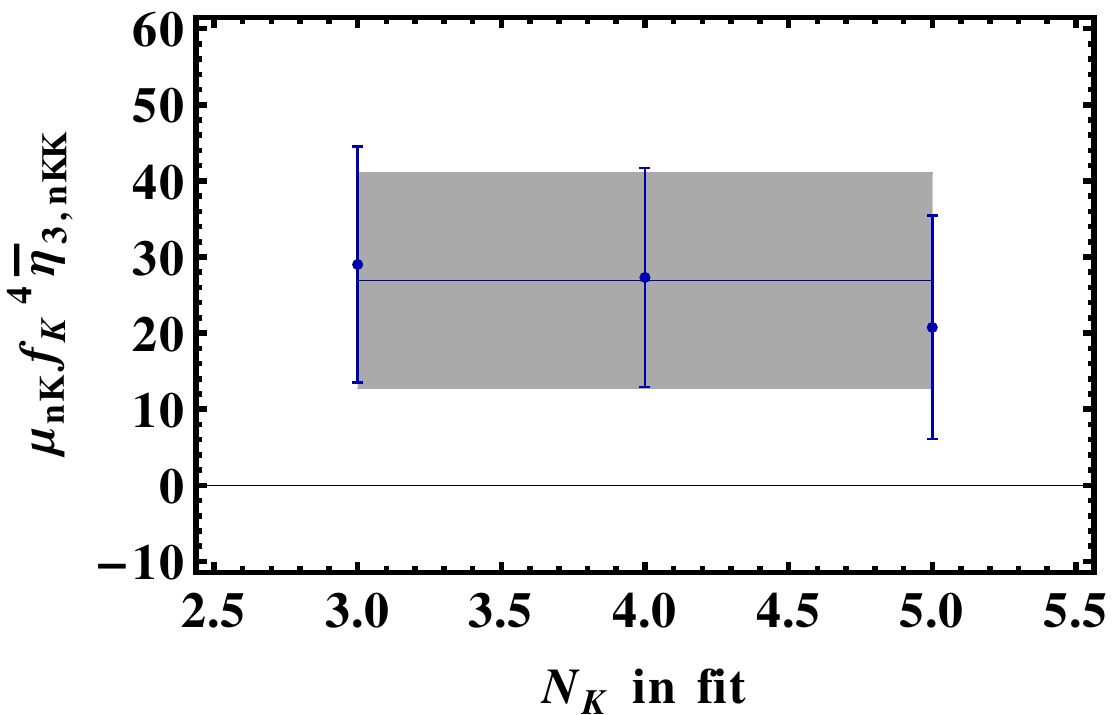}
\caption{\label{fig:etaMMB}Three-body parameters, $\bar{\eta}_{MMB}$, as a function of the maximum system size included in the fit. Clockwise from upper left: $\pi^{+}\pi^{+}\Sigma^{+}$, $\pi^{+}\pi^{+}\Xi^0$, $K^{+}K^{+}$n, and $K^{+}K^{+}$p. Error bars represent the statistical and systematic uncertainties from the individual fits combined in quadrature. The gray band shows the mean of all fits and their uncertainties, plus an additional uncertainty given by the standard deviation of all fits, added in quadrature.}
\end{figure}

In \Fig{etaMMB}, we plot the meson-meson-baryon three-body interactions extracted from our ground state energies. These are novel results, and we find nonzero contributions for most systems within our uncertainties.

\subsection{\label{sec:ChiPT}Comparison to LO $\chi$PT and extraction of LECs}

\begin{figure}
\includegraphics[width=0.48\linewidth]{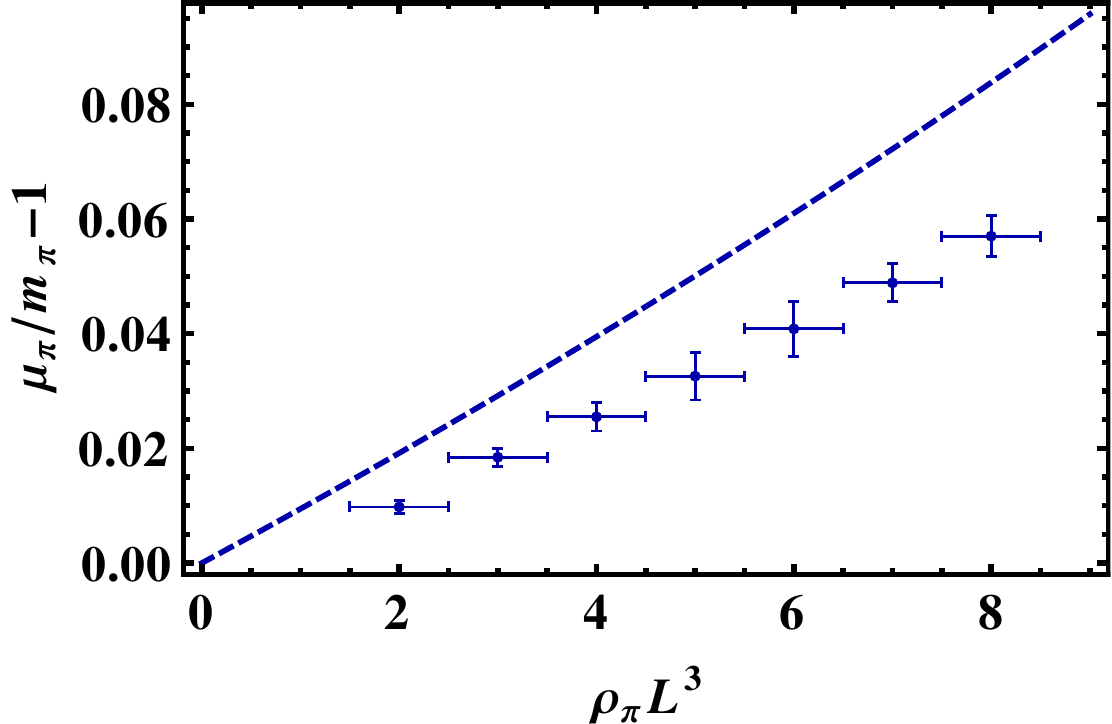}\hspace{1mm}
\includegraphics[width=0.48\linewidth]{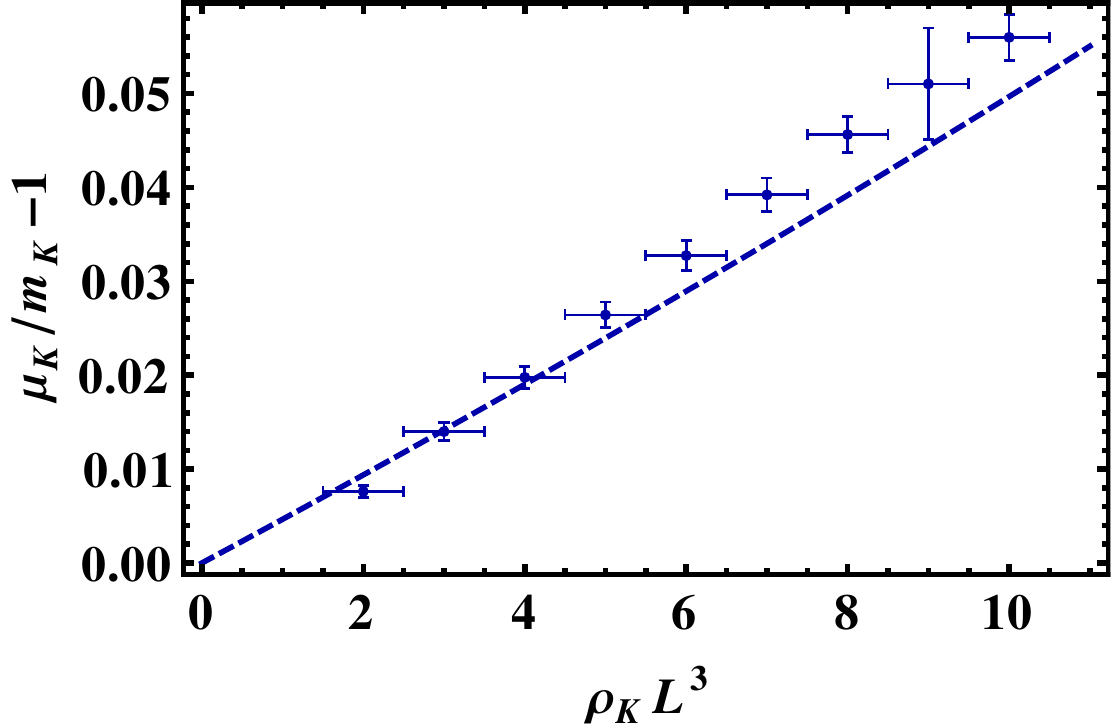}
\caption{\label{fig:denschem}Chemical potential for pions (left) and kaons (right), as a function of the number of mesons in the system. Data points were calculated using a finite energy difference from the lattice data, \Eq{muEn}, while the dashed line is the tree-level prediction from $\chi$PT, \Eq{rhovmu}.}
\end{figure}

We may extract combinations of LECs from $\chi$PT using the expressions for the baryon masses as a function of chemical potential presented in \Eqs{Sigmass}{Protmass}. To translate between the canonical and grand canonical formulations, we calculate the effective chemical potential as a finite energy difference,
\beq
\label{eq:muEn}
\mu_{I,K}(n) = E_{\pi,K}(n+1)-E_{\pi,K}(n) \ ,
\eeq
where $E_{\pi,K}(n)$ is the energy of the system of $n$ pions or kaons, respectively, and $\mu_{I,K}$ is the isospin or kaon chemical potential, respectively. This relation should be valid at low temperature in the thermodynamic limit. The behaviors of $\mu_{I,K}$ vs. $N_{I,K} = \rho_{I,K}L^3$, where $\rho_{I,K}$ is the isospin/kaon density, for the spatial volume used in this calculation are shown in Fig.~\ref{fig:denschem}. We also compare our results to the tree level prediction from $\chi$PT (\cite{Son:2000xc}),
\beq
\label{eq:rhovmu}
\rho_{I,K} = -\frac{\partial \mathcal{L}_{stat}}{\partial \mu_{I,K}} = f_{\pi,K}^2 \mu_{\pi,K} \left( 1-\frac{m_{\pi,K}^4}{\mu_{\pi,K}^4}\right) \, ,
\eeq
where $f_{\pi,K}$ and $m_{\pi,K}$ are the pion or kaon decay constant and mass, respectively. These are shown as dashed lines in the figures. 

Because the mass relations were computed using a combination of canonical and grand canonical methods, we must further modify them before relating them to our lattice data. In particular, in Ref.~\cite{Bedaque:2009yh} the baryon is treated as an external source, while the pions are produced using a chemical potential. However, the baryon itself carries isospin charge and couples to the chemical potential, thus its mass is altered even in the absence of pions. As our data only gives us access to changes in the mass due to interactions with pions, we must subtract off this direct coupling to chemical potential in vacuum. Thus for our analysis we use,

\beq
\label{eq:MassSplit}
\Delta M_{B}(\mu_{I,K}) &\equiv&  M_{B}\left(\mu_{I,K},\cos \alpha =\frac{m_{\pi,K}^2}{\mu_{I,K}^2}\right)-M_{B}(\mu_{I,K},\cos \alpha = 1) \ ,
\eeq
for a given baryon, $B$.

Finally, we note that the chemical potentials for our systems are very near the condensation point, $\mu_{I,K} \approx m_{\pi,K}$. We find that expanding the mass relations, \Eq{MassSplit}, around this point gives more reliable, stable fits for the resulting linear combinations of LECs at leading order in the expansion. The form of the fitting function that we use is given by
\beq
\label{eq:MassExp}
\Delta M_B/m_{\pi,K} = \left( 2Q_{I,K}+4a^{(B)}\right) \left( \frac{m_{\pi,K}}{\mu_{I,K}}-1\right) + \left(-Q_{I,K}+2 b^{(B)}\right) \left( \frac{m_{\pi,K}}{\mu_{I,K}}-1\right)^2 + \cdots \ ,
\eeq
where $Q_{I,K}$ is the isospin/kaon charge of the baryon. The combinations of LECs corresponding to the parameters $a^{(B)},b^{(B)}$ in the equation above are given in \Tab{ChiPT}, along with our extracted values.

\begin{table}
\centering
\caption{\label{tab:ChiPT}Linear combinations of LECs corresponding to the fit parameters $a$ and $b$ (\Eq{MassExp}), along with results from 1- and 2-parameter fits to the data. Uncertainties include statistical, fitting systematic, and the standard deviation from all fits to different system sizes.}
\begin{tabular}{|c|c|c|c|}
\hline
& LEC combination$/m_{\pi,K}$ & 1-param fit & 2-param fit\\
\hline
$ a^{\Sigma}$ &$c_2^{\Sigma}+c_3^{\Sigma}+c_6^{\Sigma}+c_7^{\Sigma}-2 c_1^{\Sigma}$&0.80(20)&0.86(28)\\
$ b^{\Sigma}$  &$6 c_1^{\Sigma}-c_2^{\Sigma}-c_3^{\Sigma}-c_6^{\Sigma}-c_7^{\Sigma}-4\left(c_6^{\Sigma}+c_7^{\Sigma}\right)^2m_{\pi}$&-&-4(14) \\
\hline
 $a^{\Xi} $& $c_2^{\Xi}-\frac{g_{\Xi}}{8M_{\Xi}^{(0)}}+c_3^{\Xi}-2c_1^{\Xi}$&0.29(14)&0.26(22) \\
$ b^{\Xi}$ &$6c_1^{\Xi}- c_2^{\Xi}+\frac{g_{\Xi}}{8M_{\Xi}^{(0)}}-c_3^{\Xi}$&-&4(18) \\
\hline
$ a^p$ &$\frac{1}{2}\left(b_1+b_3+b_4+b_6+b_7+b_8\right)-b_0-b_D$ &0.76(14)&0.86(24) \\
  \multirow{2}{*}{$b^p$} &$3\left(b_0+b_D\right)-\frac{1}{2}\left(b_1+b_3+b_4+b_6+b_7+b_8-1\right)$&\multirow{2}{*}{-}&\multirow{2}{*}{-7(13)}\\
&$+\frac{-1+2\left(b_1+b_3-b_4+b_6\right)m_K^2-2b_F/m_K\left(m_K^2-m_{\pi}^2\right)}{m_{K}+2b_F\left(m_{K}^2-m_{\pi}^2\right)}$&& \\
\hline
 $a^n$ &$\frac{1}{4}\left(b_1-b_2+b_3+b_4-b_5+b_6\right)-b_0+\frac{1}{2}\left(b_F-b_D+b_7+b_8\right)$&0.25(11)& 0.35(33)\\
 $b^n$ &$3b_0-\frac{1}{4}\left(b_1-b_2+b_3+b_4-b_5+b_6+2b_7+2b_8\right)+\frac{3}{2}\left(b_D-b_F\right)$&-&-10(33)\\
\hline
\end{tabular}
\end{table}

\begin{figure}
\includegraphics[width=0.48\linewidth]{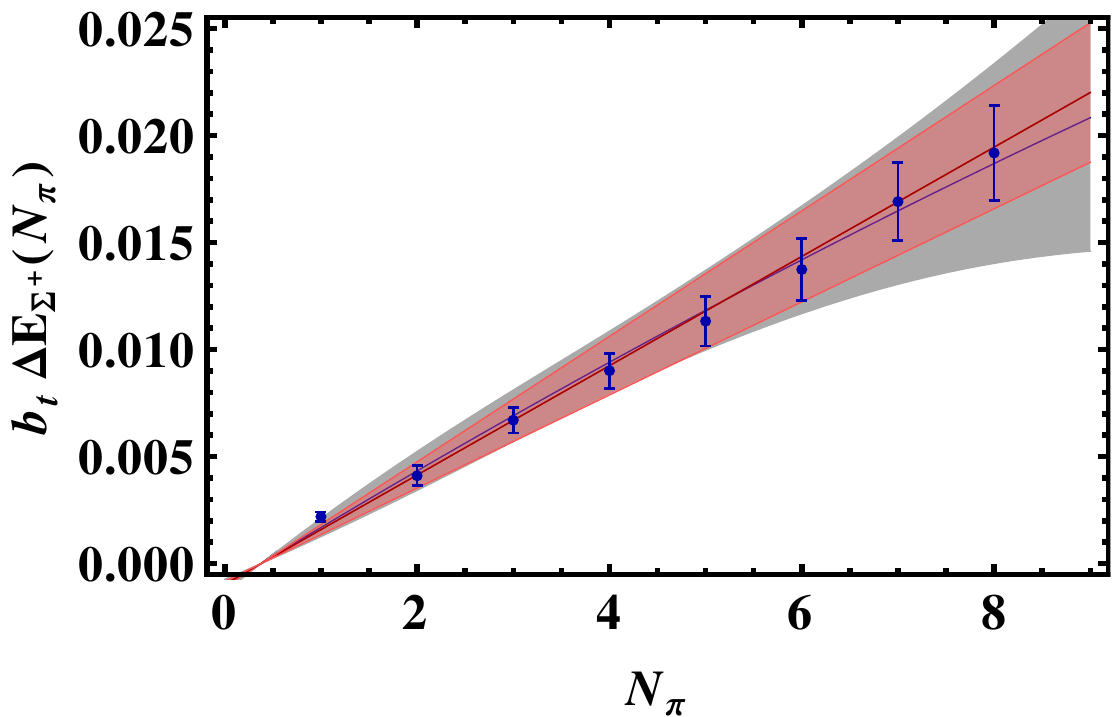}\hspace{1mm}
\includegraphics[width=0.48\linewidth]{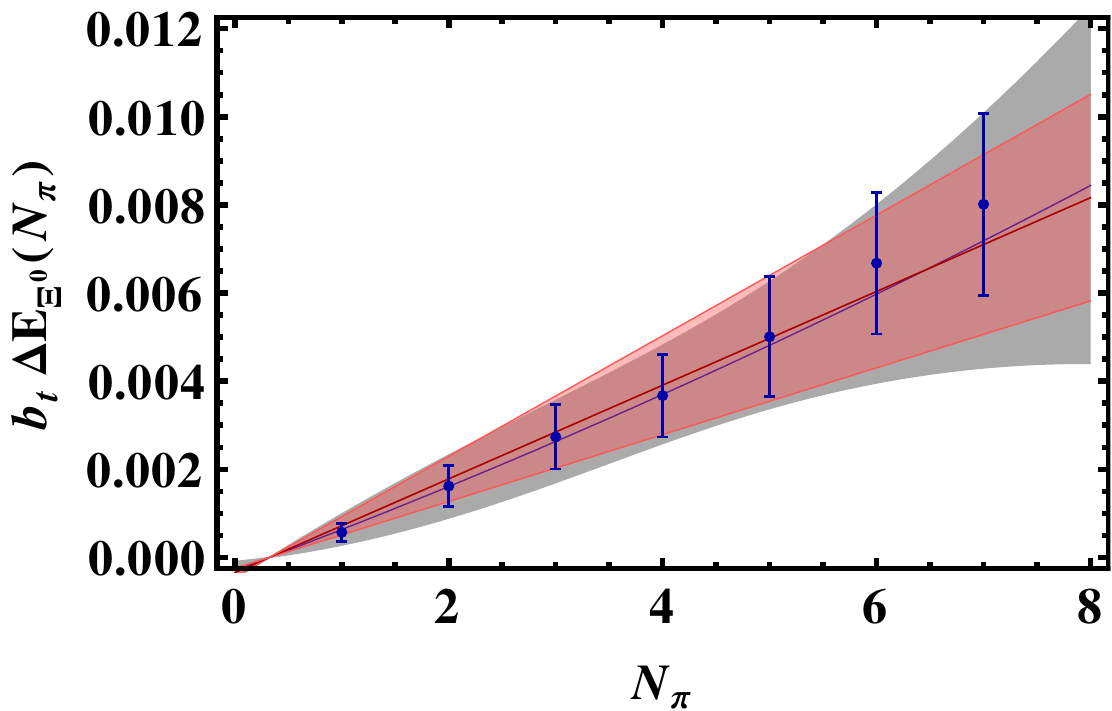} \\

\vspace{2mm}

\includegraphics[width=0.48\linewidth]{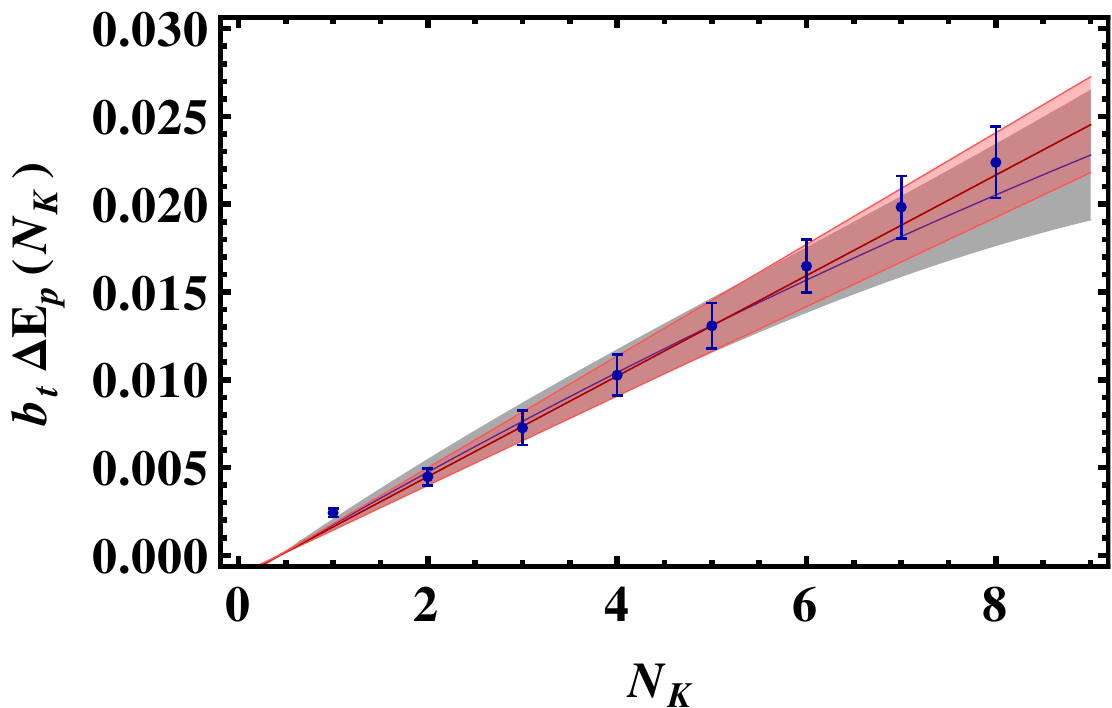} \hspace{1mm}
\includegraphics[width=0.48\linewidth]{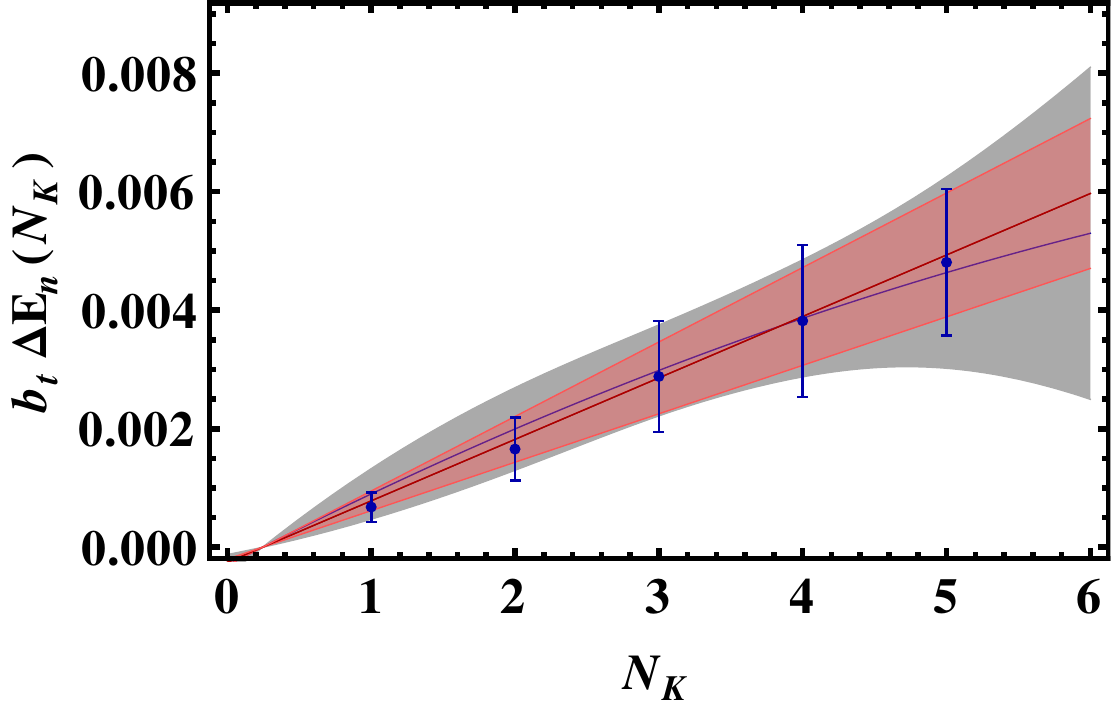}
\caption[]{\label{fig:ChiPT}Fit results for the energy splittings of the meson-baryon systems as a function of the number of mesons. Clockwise from upper left:  $(\pi^{+})^{N_{\pi}}\Sigma^{+}$, $(\pi^{+})^{N_{\pi}}\Xi^0$, $(K^{+})^{N_K}$n, and $(K^{+})^{N_K}$p. The error bars represent the statistical and fitting systematic uncertainties combined in quadrature. One- and two-parameter fits to the $\chi$PT form for the mass splittings, \Eq{MassExp}, are shown as red and gray shaded bands, respectively.}
\end{figure}

The ground-state energy shifts of all systems as a function of the number of mesons in the system are shown in \Fig{ChiPT}, along with the results of a fit to the form in \Eq{MassExp}. Both one- and two-parameter fits have been performed, shown in red and gray, respectively. We see little improvement to the fit by including the second term. 

\begin{figure}
\includegraphics[width=0.48\linewidth]{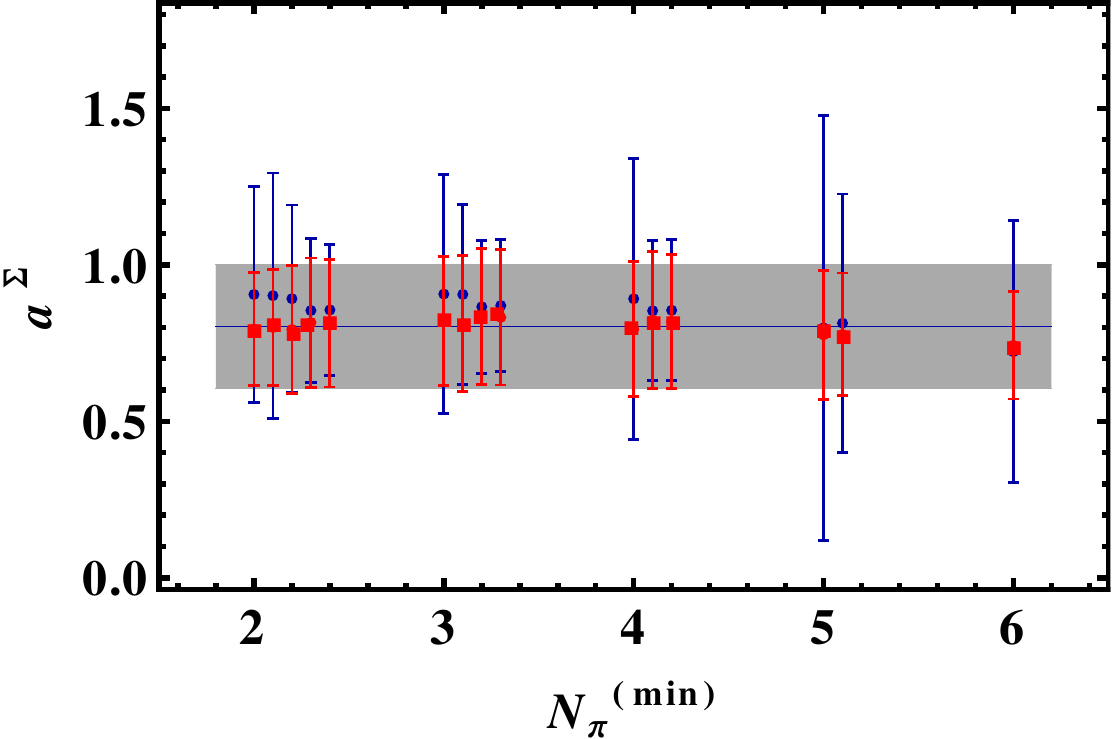}
\includegraphics[width=0.48\linewidth]{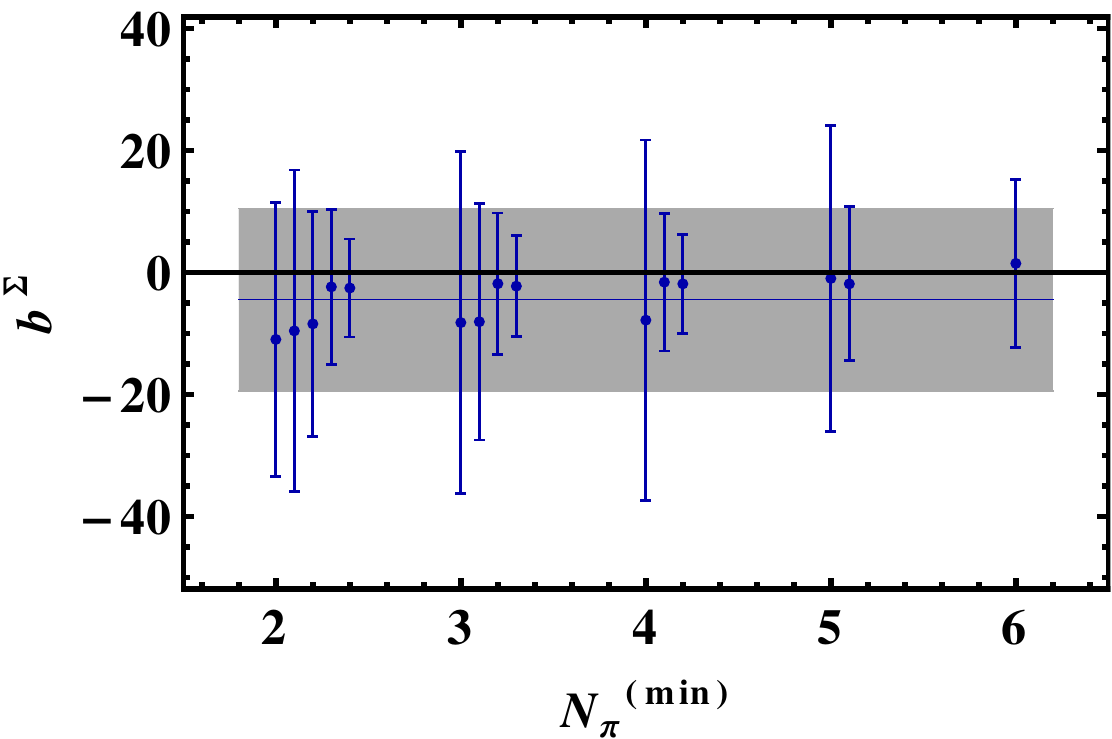}
\caption{\label{fig:abSig}One- and two-parameter fit results (red squares and blue circles, respectively) for $a^{\Sigma}$ and $b^{\Sigma}$ from the expansion of the $\chi$PT prediction for the mass splitting of the $\Sigma^{+}$ + $N_{\pi}$-pions systems, Eq.~(\ref{eq:MassExp}), as a function of the minimum system size included in the fit. Each cluster of points includes higher maximum system sizes from left to right. Error bars represent the statistical and systematic uncertainties from the individual fits combined in quadrature. The gray band shows the mean of all 1- and 2- parameter fits for $a^{\Sigma}$ and $b^{\Sigma}$, respectively, and their uncertainties, plus an additional uncertainty given by the standard deviation of all fits, added in quadrature.}
\end{figure}

The extracted parameters $a^{(B)},b^{(B)}$ are shown in \Figs{abSig}{abNeut} as a function of the minimum number of mesons included in the fit, with clusters of points representing different maximum numbers of mesons. The results for one and two parameter fits are shown in red and blue, respectively. For the $a^{(B)}$ parameter in each case we see no change by including the second parameter in the fit. We are unable to resolve the $b^{(B)}$ parameters within our uncertainties. Thus, we are only able to extract the linear combinations of LECs corresponding to the $a^{(B)}$ parameters shown in column 3 of \Tab{ChiPT}. To further isolate the individual LECs would require either larger chemical potentials or variation of the quark masses.

\begin{figure}
\includegraphics[width=0.48\linewidth]{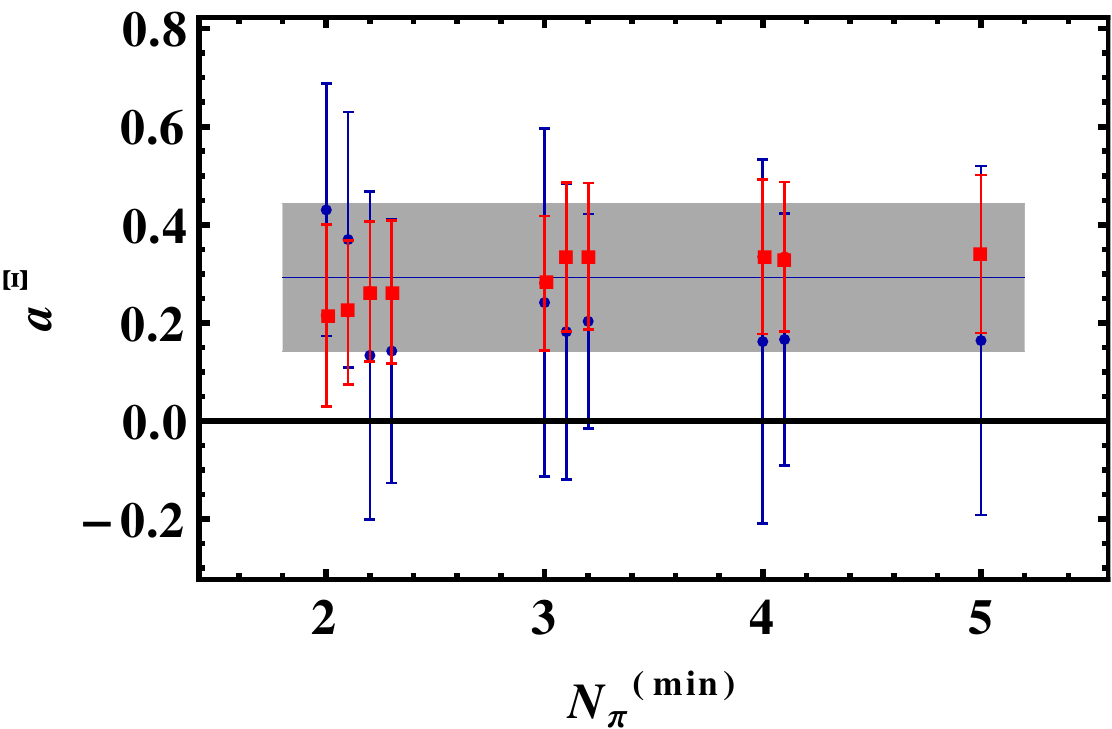}\hspace{1mm}
\includegraphics[width=0.48\linewidth]{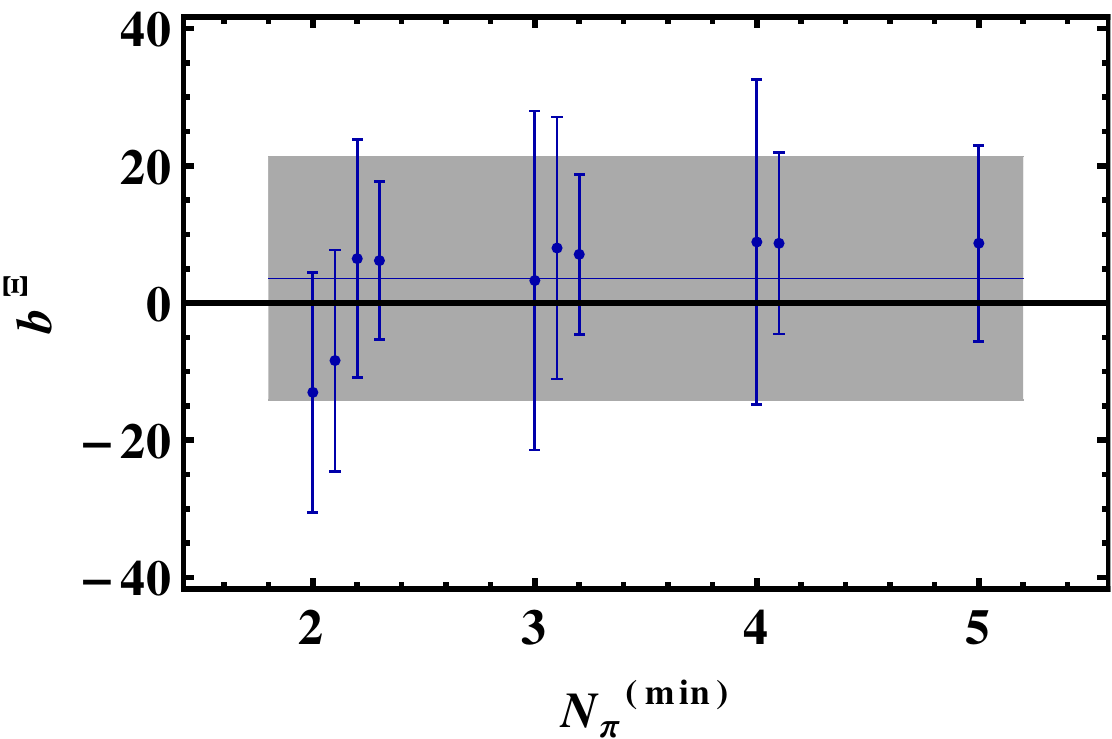}
\caption{\label{fig:abXi}One- and two-parameter fit results (red squares and blue circles, respectively) for $a^{\Xi}$ and $b^{\Xi}$ from the expansion of the $\chi$PT prediction for the mass splitting of the $\Xi^{0}$ + $N_{\pi}$-pions systems, Eq.~(\ref{eq:MassExp}), as a function of the minimum system size included in the fit. Each cluster of points includes higher maximum system sizes from left to right. Error bars represent the statistical and systematic uncertainties from the individual fits combined in quadrature. The gray band shows the mean of all 1- and 2- parameter fits for $a^{\Xi}$ and $b^{\Xi}$, respectively, and their uncertainties, plus an additional uncertainty given by the standard deviation of all fits, added in quadrature.}
\end{figure}

\begin{figure}
\includegraphics[width=0.48\linewidth]{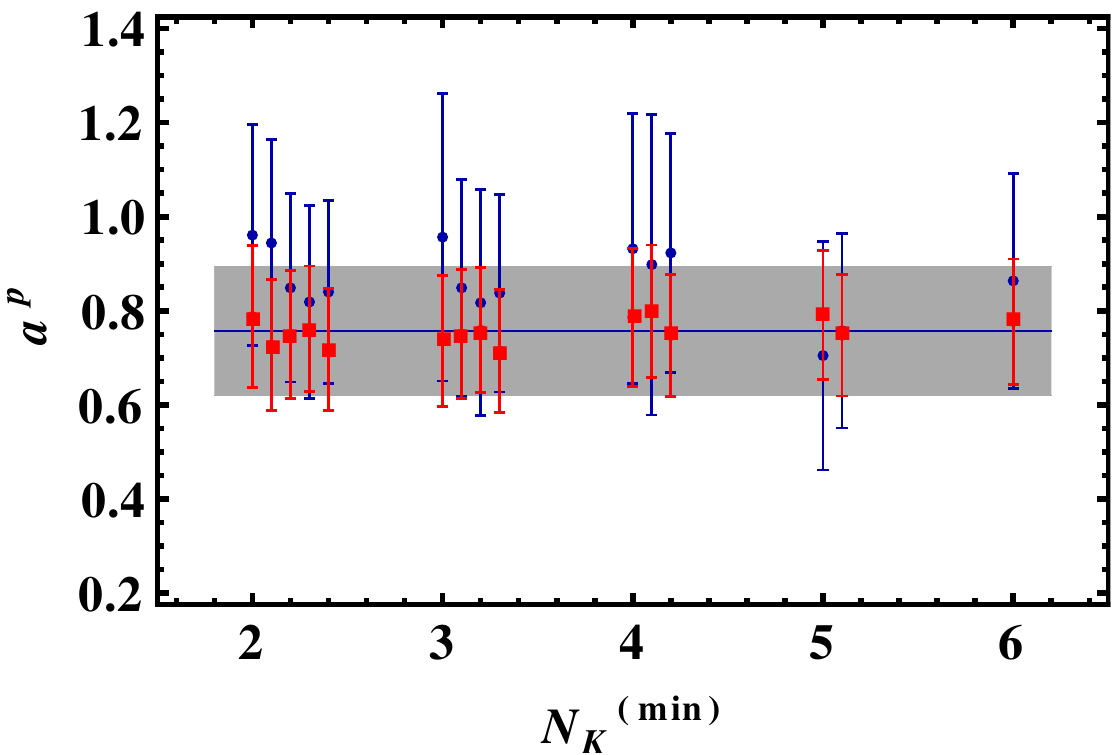}\hspace{1mm}
\includegraphics[width=0.48\linewidth]{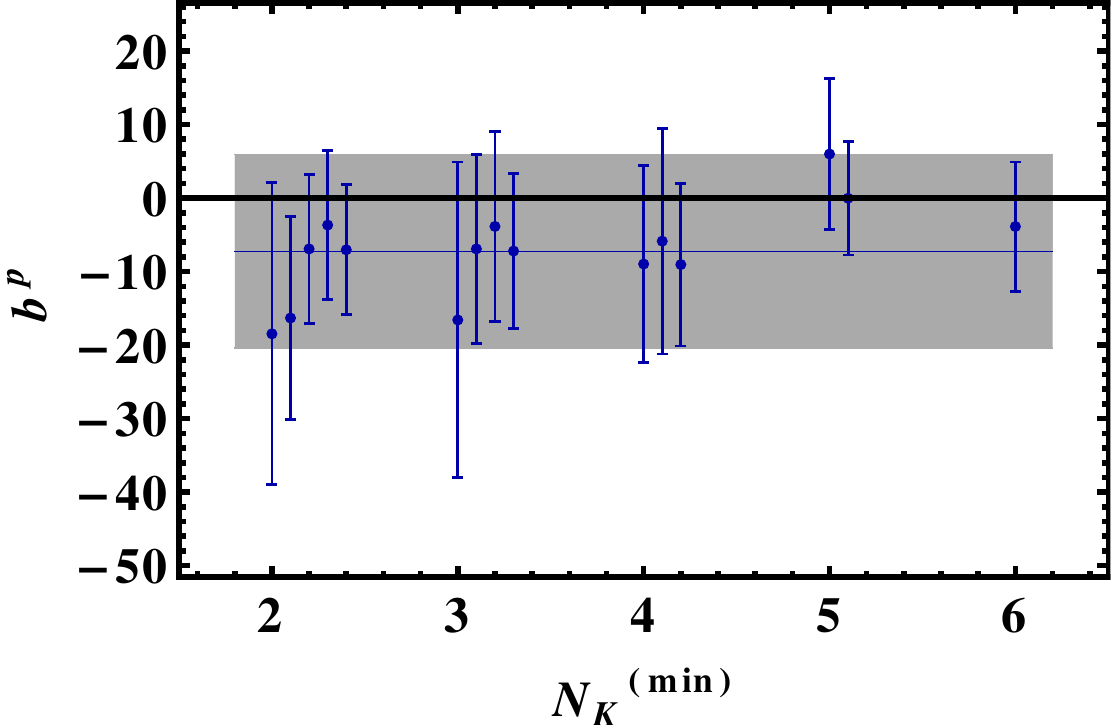}
\caption{\label{fig:abProt}One- and two-parameter fit results (red squares and blue circles, respectively) for $a^{p}$ and $b^{p}$ from the expansion of the $\chi$PT prediction for the mass splitting of the proton + $N_{K}$-kaons systems, Eq.~(\ref{eq:MassExp}), as a function of the minimum system size included in the fit. Each cluster of points includes higher maximum system sizes from left to right. Error bars represent the statistical and systematic uncertainties from the individual fits combined in quadrature.  The gray band shows the mean of all 1- and 2- parameter fits for $a^{p}$ and $b^{p}$, respectively, and their uncertainties, plus an additional uncertainty given by the standard deviation of all fits, added in quadrature.}
\end{figure}

\begin{figure}
\includegraphics[width=0.48\linewidth]{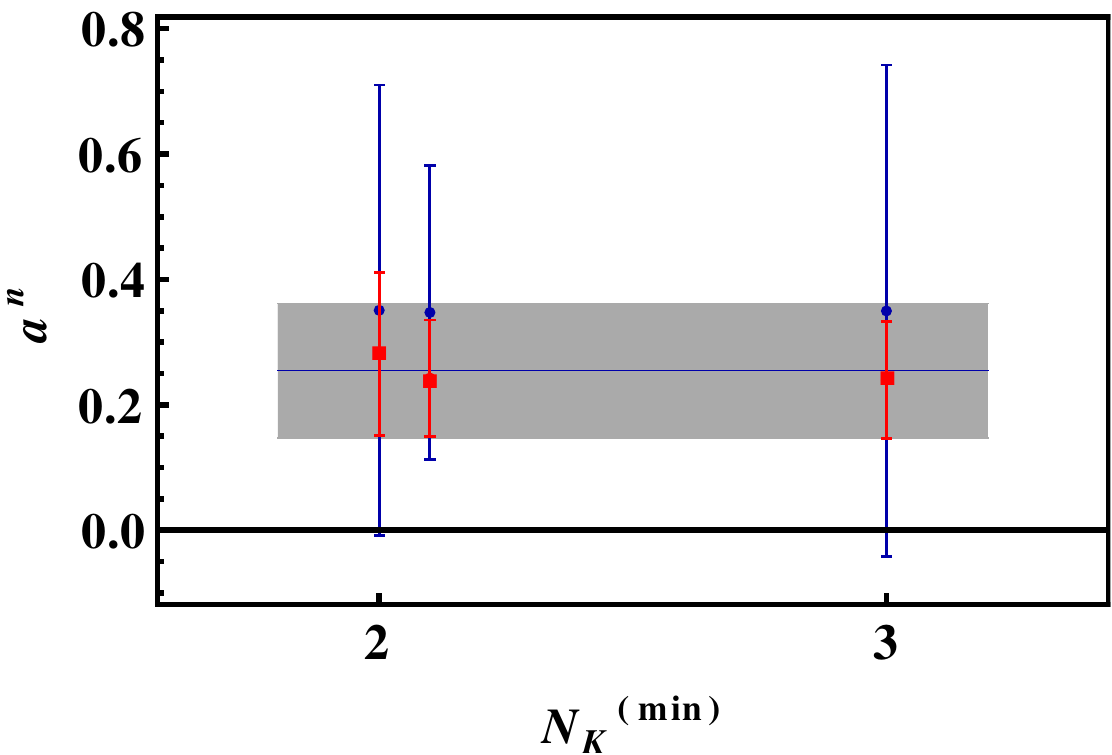}\hspace{1mm}
\includegraphics[width=0.48\linewidth]{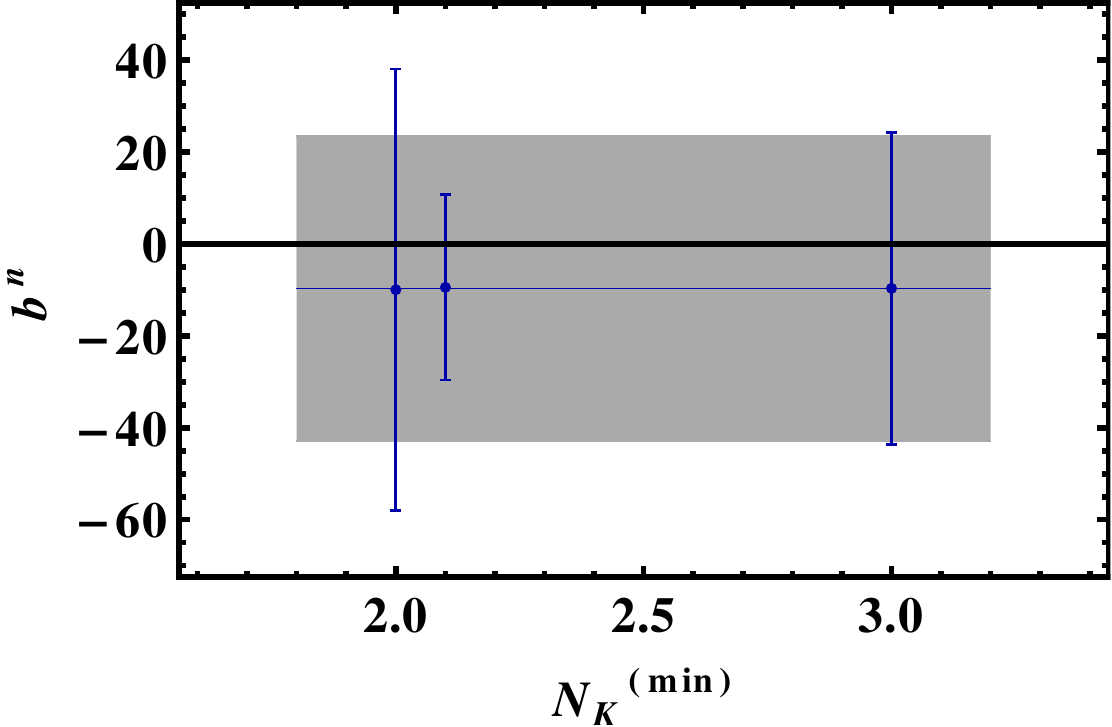}
\caption{\label{fig:abNeut}One- and two-parameter fit results (red squares and blue circles, respectively) for $a^{n}$ and $b^{n}$ from the expansion of the $\chi$PT prediction for the mass splitting of the neutron + $N_{K}$-kaons systems, Eq.~(\ref{eq:MassExp}), as a function of the minimum system size included in the fit. Each cluster of points includes higher maximum system sizes from left to right. Error bars represent the statistical and systematic uncertainties from the individual fits combined in quadrature.  The gray band shows the mean of all 1- and 2- parameter fits for $a^{n}$ and $b^{n}$, respectively, and their uncertainties, plus an additional uncertainty given by the standard deviation of all fits, added in quadrature.}
\end{figure}

\section{Conclusions}

In this study we have calculated the energy shifts of single baryons in the presence of a meson medium using lattice QCD. We have presented results for up to 9 mesons, and have extracted two- and three-body interactions from the energy splittings. By comparing our results with previous ones from NPLQCD \cite{Torok:2009dg}, we find that the meson-baryon scattering phase shifts have nontrivial momentum-dependence at momenta much smaller than the pion mass. We are able to resolve non-zero three-body interactions for most systems within our uncertainties. We have also compared to tree level $\chi$PT results, and extracted certain linear combinations of LEC's from our data. 

While these results represent a first step toward determining the response of baryons to the presence of a meson medium, to make a connection to the physical limit we need to explore both the pion mass and lattice spacing dependence of these quantities. To connect with phenomenologically relevant systems we must include multiple baryons, as well as explore channels involving disconnected diagrams. Both of these pursuits would require considerably larger computational resources than those used for this study, so we leave these systems for future exploration. Finally, it would also be enlightening to relax the restriction of zero momentum in the baryon systems, to explore the non-trivial dispersion relations predicted by \cite{Birse:2001sn}. 

\begin{acknowledgments}
The authors would like to thank A. Walker-Loud, R, Brice\~{n}o, B. Smigielski, J. Wasem, P. Bedaque, S. Wallace, M. Savage, and D. Kaplan for helpful discussions. We thank the NPLQCD collaboration for allowing us to use previously generated meson and baryon blocks. WD was supported in part by the U.S. Department of Energy through Outstanding Junior Investigator Award DE-SC000-1784 and Early Career Award DE-SC0010495. AN was supported in part by U.S. DOE grant No. DE-FG02-93ER-40762. 

\end{acknowledgments}
\bibliography{Mesonsref}

\end{document}